\documentclass[a4paper,10pt,twoside]{cpc-hepnp}
\usepackage{multicol}
\usepackage{graphicx}
\usepackage{booktabs}
\usepackage{amssymb,bm,mathrsfs,bbm,amscd}
\usepackage[tbtags]{amsmath}
\usepackage{lastpage} 
\usepackage{slashed}  
\usepackage{url,hyperref}  
\usepackage{tabularx}  
\usepackage{multirow}  
\usepackage{threeparttable}  
\usepackage{float}  
\usepackage{subfig}  
\usepackage{color}  
\usepackage{caption}  
\usepackage{fancyhdr}  
\usepackage{placeins}  
\usepackage{footmisc}

\definecolor{darkred}{rgb}{.8,0,0}

\definecolor{darkblue}{rgb}{0,0,.7}

\definecolor{darkgreen}{rgb}{0,.7,0}

\begin{document}

\fancyhead[c]{\small Chinese Physics C~~~Vol. xx, No. xx (20xx) 010201}
\fancyfoot[C]{\small xxxxxx-\thepage}

\footnotetext[0]{Received xx Feb 2025}

\title{Investigating the shadows of new regular black holes with a Minkowski core: Effects of spherical accretion and core type differences}

\author{%
      Yi Xiong $^{1}$%
\quad Pu Jin $^{1;1)}$\email{pujin@cwnu.edu.cn}%
\quad Yi Ling $^{2,3}$ %
\quad Guo-Ping Li $^{1}$%
\quad Gao-Ming Deng $^{1}$
}

\maketitle

\address{%
    $^1$ School of Physics and Astronomy, China West Normal University, Nanchong 637002, China\\
    $^2$ Institute of High Energy Physics, Chinese Academy of Sciences, Beijing 100049, China\\
    $^3$ School of Physics, University of Chinese Academy of Sciences, Beijing 100049, China
}

\begin{abstract}
We investigated the shadows and optical appearances of a new type of regular black holes (BHs) with a Minkowski core under various spherical accretion scenarios. These BHs are constructed by modifying the Newtonian potential based on the minimum observable length in the Generalized Uncertainty Principle (GUP). They correspond one-to-one with traditional regular BHs featuring a de-Sitter (dS) core (such as Bardeen/Hayward BHs), characterized by a quantum gravity effect parameter ($\alpha_0$) and spacetime deformation factor ($n$). We found that the characteristic parameters give rise to some novel observable features. For these new BHs, both the shadow and photon sphere radii decrease with the increase in $\alpha_0$, while the observed specific intensity increases. Conversely, as n increases, the shadow and photon sphere radii increase, while the observed specific intensity decreases. Under different spherical accretion scenarios, the shadows and photon sphere radii remain identical; however, the observed specific intensity is greater under static spherical accretion than under infalling spherical accretion. Additionally, we found that these regular BHs with different cores exhibit variations in shadows and optical appearances, particularly under static spherical accretion. Compared with Bardeen BH, the new BHs exhibit a lower observed specific intensity, a dimmer photon ring, and smaller shadow and photon sphere radii. Larger values of $\alpha_0$ lead to more significant differences, and a similar trend was also observed when comparing with Hayward BH. Under infalling spherical accretion, the regular BHs with different cores exhibit only slight differences in observed specific intensity, which become more evident when $\alpha_0$ is relatively large. This suggests that the unique spacetime features of these regular BHs with different cores can be distinguished through astronomical observation.

\end{abstract}



\footnotetext[0]{\hspace*{-3mm}\raisebox{0.3ex}{$\scriptstyle\copyright$}2025
Chinese Physical Society and the Institute of High Energy Physics
of the Chinese Academy of Sciences and the Institute
of Modern Physics of the Chinese Academy of Sciences and IOP Publishing Ltd}%

\begin{multicols}{2}

\section{Introduction}\label{intro}

The prediction of BHs by General Relativity (GR) has always been a hot topic of scientific research, sparking widespread attention and academic debate. After the Laser Interferometer Gravitational-Wave Observatory (LIGO) first detected gravitational waves from the merger of binary BHs \cite{1,2}, LIGO and its partners have recorded nearly a hundred such binary events. These observations not only validate GR but also provide rich data for BH research. Later, the Event Horizon Telescope (EHT) collaboration published the first ultra-high angular resolution image of the supermassive BH in the Messier 87 (M87*) galaxy, marking a significant milestone in astrophysical research \cite{3}. The image revealed a bright ring-like structure surrounding the central dark region, known as the BH shadow, which is formed by light being strongly bent and focused as it approaches the BH's edge. This ring-like structure is referred to as the photon ring. The strong gravitational field around the BH is a crucial factor in the formation of the shadow. Furthermore, EHT's polarization measurements have revealed the significant impact of the strong magnetic field at the edge of the BH on the behavior of surrounding matter, confirming that the magnetic field structure around the M87* BH's accretion disk aligns with the predictions of general relativistic magnetohydrodynamics (GRMHD) models \cite{4}. These findings contribute to our understanding of the physical processes around BHs, including the infall of matter into BHs and the formation of BH jets. Afterwards, EHT captured the first hierarchical radio observation image of the Sagittarius A* (SgrA*) BH at the center of our Milky Way galaxy, which is mainly composed of a bright thick ring \cite{5}. These observations suggest that the accretion disk may influence the observational characteristics of the BH shadows, offering a new perspective for studying BH shadows.

Indeed, the theoretical analysis of the shadow formation of Schwarzschild BH was proposed even before EHT \cite{6}. Here, the escaping cone of photon was defined by the impact parameter corresponding to the critical curve, which is known as the BH shadow. Later, Bardeen examined the D-shaped shadow of the Kerr BH, which arises from the dragging effect of the rotating BH on the path of light rays \cite{7}. Additionally, the actual astrophysical BHs in the universe are always encircled by various accretion materials, which are of paramount importance for the observation and imaging of BHs. Consequently, Shakura and Sunyaev postulated a standard thin accretion disk model, assuming the accretion disks are geometrically thin, optically thick, and in a steady state \cite{8}. This model predicts the disk's geometric configuration, optical properties, and spectral characteristics, providing a theoretical foundation for understanding the fundamental nature and energy release of accretion disks. This model is also in alignment with numerous BH observational phenomena. Using this thin accretion disk model, Luminet simulated the first image of a BH with an emitting accretion disk using a semi-analytic method, visually illustrating the gravitational lensing effect and the formation of the photon ring \cite{9}.

However, to more closely approximate the accretion flows around BHs in the realistic astrophysical environment, it is necessary to consider the interaction of magnetic fields, which requires numerical simulations using GRMHD to extract the BH images. It is worth noting that in most cases, in order to reveal the basic characteristics of BH images and the behavior of strong gravitational fields, using some toy models of accretion structures is sufficient for research. In this sense, Wald et al. studied the radiation from optically and geometrically thin disk accretion around Schwarzschild BHs, categorizing it into direct, lensed ring, and photon ring emissions \cite{10}. They found that the observed BH images are primarily dominated by the direct emissions, with smaller contributions from lensed ring emissions and negligible contributions from photon ring emissions. The observed appearance of the BH shadow depends on the emission details and morphology of the disk accretion, including the geometry of the emission region, the emission profile, and the optical depth. This initiated a series of research efforts on various toy accretion models for different types of BHs \cite{11,12,13,14,15,16,17,18,19,20,21}. Spherical accretion, as a simplified toy accretion model, has also garnered widespread attention \cite{22,23,24,25,26,27,28}. In this model, the size and shape of the BH shadow are mainly determined by the spacetime geometry of the BH, rather than the specific details of the accretion process. Thus, a large number of simulations of various BH shadows have been widely discussed, as shown in examples and references therein \cite{29,30,31,32,33,34,35,36,37,38,39,40,41,42,43,44,45,46}. These studies indicate that the properties of BH shadows are closely related to the spacetime background, which also provides a potential method for distinguishing BHs in GR from other alternative compact objects or BHs in alternative theories beyond GR.

To avoid the infamous singularity problem, researchers have proposed constructing regular BHs without singularities  phenomenologically, either by introducing exotic matter that violates standard energy conditions or through quantum corrections to spacetime geometry \cite{47,48,49,50,51,52,53,54,55,56,57}. Traditional regular BHs, such as Bardeen/Hayward/Frolov BHs, have a dS core at their centers. Recently, a new type of regular BHs with sub-Planckian curvature was introduced in \cite{58,59}, which is characterized by an exponentially suppressed Newtonian potential and an asymptotically Minkowski core. These new regular BHs can reproduce the metric of Bardeen/Hayward/Frolov BHs on large scales by choosing different forms of the potential, and is derived from GUP in curved spacetime. Such regular BHs with a Minkowski core have also been reported in \cite{60,61}, but our understanding of them remains incomplete. Hence, exploring the connection between these new regular BHs and astronomical observations to search for any signs or traits that might indicate that the BHs observed in the cosmos belong to the regular type rather than the traditional singular BHs, has become an intriguing endeavor.

Evidently, at this juncture, it is impractical to differentiate between singular BHs and regular BHs by diving into the event horizon or detecting any signals from their interiors. Nevertheless, the distinctions within the BHs might manifest in external phenomena, such as the photon sphere or the trajectories of massive particles \cite{62,63,64,65,66,67,68,69,70,71,72,73,74}.Therefore, in \cite{62}, we meticulously examined the photon spheres and marginally stable circular orbits for regular BHs with two distinct cores: these new regular BHs with a Minkowski core, and Bardeen/Hayward BHs with a dS core. The results show that the positions of the photon sphere and the marginally stable circular orbits for the compact massive object (CMO) phase with a Minkowski core are notably distinct from their counterparts in the CMO phase with a dS core. Following this, in \cite{15}, we investigated the thin accretion disks of regular BHs with a Minkowski core that can reproduce Bardeen BH at large scales, and compared them with the accretion disks around Bardeen BH with a dS core. The results reveal that the thin accretion disks of these regular BHs with different cores exhibit distinct astronomical optical characteristics.

On this basis, we have extended the work presented in \cite{15} by studying images of these new regular BHs with a Minkowski core under the spherical accretion, and comparing them with those of traditional regular BHs with a dS core. This provides the theoretical foundation for distinguishing between the two types of regular BHs with different cores through astronomical observations. More importantly, in \cite{62}, we found that for the regular BHs with a Minkowski core, different choices of the potential form result in different positions of the photon spheres and marginally stable circular orbits in the CMO phase. Additionally, it was observed in \cite{15} that the optical characteristics of Bardeen and Hayward BHs, surrounded by thin accretion disks, exhibit distinct differences. In particular, the Hayward BH exhibits a minimal distance between the innermost region of the direct image and the outermost region of the secondary image, which is anticipated to serve as an identifier for distinguishing the Hayward BH. These findings have sparked our interest in exploring whether there are also distinct differences in the images of these new regular BHs with a Minkowski core under the spherical accretion model, which would be a very meaningful pursuit.

The paper is organized as follows. In Sec.\ref{sec2}, we briefly review these new regular BHs with a Minkowski core, which can reproduce the characteristics of traditional regular BHs on large scales, and compare the behavior of light rays around these regular BHs with different cores by using a ray-tracing code. In Sec.\ref{sec3}, we will study the shadow and optical appearances of these new regular BHs with a Minkowski core under two types of spherical accretion. In Sec.\ref{sec4}, we will compare the shadows of the regular BHs with a Minkowski core with those of traditional regular BHs (Bardeen/Hayward BHs) with a dS core. Our conclusions and discussions are presented in Sec.\ref{sec5}.
\end{multicols}
\begin{multicols}{2}
\section{The new regular BHs with a Minkowski core}\label{sec2}
In order to avoid singularity, the key to constructing the regular BH at the phenomenological level is to obtain a finite value for Kretschmann scalar curvature. A new type of regular BHs with an asymptotically Minkowski core was proposed in \cite{58}, and the metric for this type of regular BHs has the following form
\begin{equation}
ds^2=-f(r)dt^2+\frac{1}{f(r)}dr^2 +r^2d\Omega^2,
\label{eq:1}
\end{equation}
with
\begin{equation}
f(r)= 1+2 \psi(r).
\label{eq:1-1}
\end{equation}
The gravitation potential $\psi(r)$ is defined as
\begin{equation}
\psi(r)=-\frac{M}{r} e^{-\alpha_{0}\frac{M^x}{r^n}}.
  \label{eq:2}
\end{equation}
The dimensionless parameter $\alpha_{0}$ is induced by quantum gravity effects \footnote{In order to ensure the exponential factor in Eq.(\ref{eq:2}) remains dimensionless throughout, we implicitly adjust via appropriate dimensional powers of the Planck length $l_p=1$. Explicitly, one may define a parameter $\alpha=\alpha_0 l_p^{n-x}$ and express $G$ into a form of $M^x$, such that the exponential term takes the form $-\alpha(GM)^x /r^n$. In this form, the parameter $\alpha$ carries the dimension $l_p^{n-x}$, thereby explicitly revealing its origin in the effects of quantum gravity.} The gravitation potential has an exponentially suppressed form, and the motivation stems from the modification of the standard Heisenberg uncertainty principle due to quantum gravity \cite{75,76,77,78,79,80}.

The dimensionless parameters $x$ and $n$ together determine the behavior of Kretschmann scalar curvature at the core, where they must satisfy the conditions $n \geq x \geq n/3$ and $n \geq 2$ in order to ensure the existence of the event horizon and maintain the scalar curvature at sub-Planckian levels. Obviously, as $r$ approaches $0$, the spacetime is dominated by a Minkowski core. Moreover, for certain values of $x$ and $n$, a one-to-one correspondence can be established between such new regular BHs with a Minkowski core and those with a dS core on large scales \cite{58}. Specifically, for the new regular BHs that have a Minkowski core, as outlined in Eq.(\ref{eq:2}), there exists the corresponding traditional regular BHs with a dS core, and their gravitation potentials are given by
\begin{equation}
\label{eq:3}
\psi(r)=-\frac{M r^{\frac{n}{x}-1}}{\left(r^{n}+x \alpha_{0} M^{x}\right)^{1 / x}}.
\end{equation}
Observe that on large scales, the new regular BH with parameters $x=2/3$ and $n=2$ corresponds to Bardeen BH, while with parameters $x=1$ and $n=3$, it corresponds to Hayward BH. This indicates that the two types of regular BHs exhibit the same asymptotic behavior on large scales, but they have distinct cores at their centers. It should be noted that for these new regular BHs with a Minkowski core, the quantum gravity effect parameter $\alpha_0$ must be within the range $0 \leq \alpha_0 \leq 0.73M$ for the case of $x=2/3$ and $n=2$, and within the range $0 \leq \alpha_0 \leq 0.98M$ for $x=1$ and $n=3$ to ensure the existence of the horizon. When $\alpha_0=0$, it returns to the standard Schwarzschild BH.

\begin{figure}[H]
    \centering
    \subfloat[]{\includegraphics[width=0.33\textwidth]{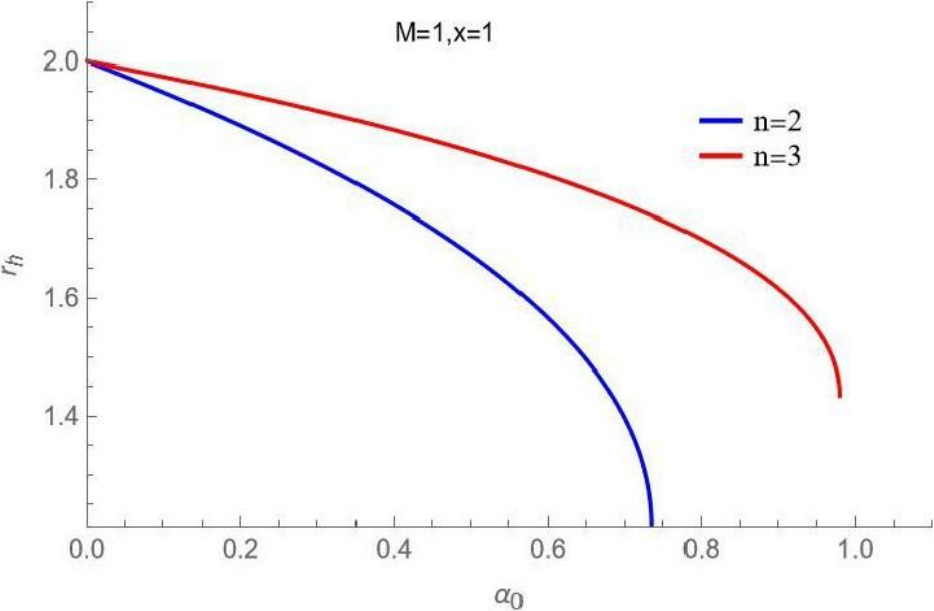}}
    \subfloat[]{\includegraphics[width=0.33\textwidth]{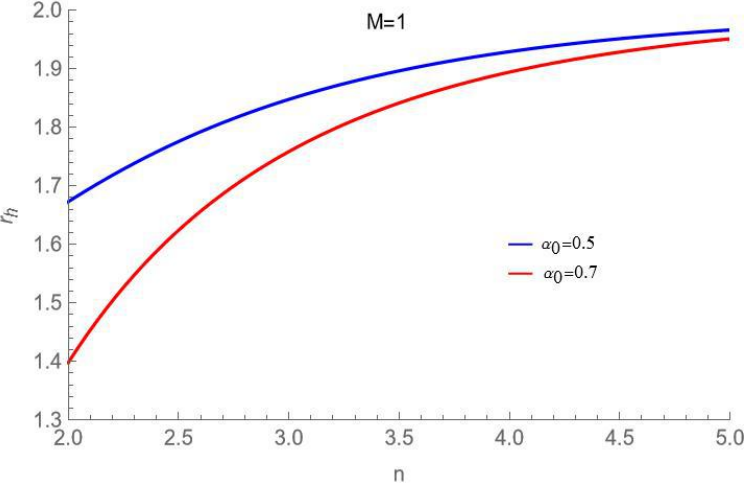}}
    \subfloat[]{\includegraphics[width=0.33\textwidth]{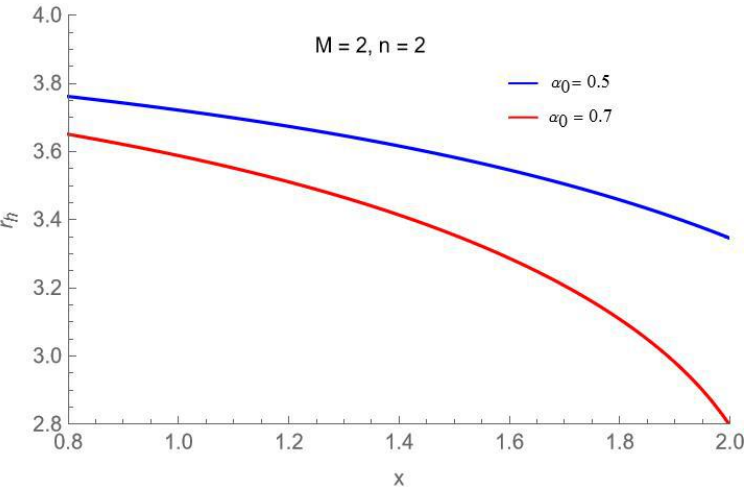}}
    \caption{ The event horizon $r_h$ of the new regular BHs as functions of the parameters $\alpha_0$, $n$ and $x$.
    \label{fig:1}}
\end{figure}

The horizon of the line element given in Eq.(\ref{eq:1}) is determined by solving $1+2 \psi(r_h)=0$. Following the numerical calculation, we illustrate the influence of the parameters $\alpha_0$, $n$ and $x$ on the event horizon $r_h$ while keeping the other parameters constant in Fig.(\ref{fig:1}). Obviously, with fixed $n$ and $x$, the event horizon $r_h$ decreases as $\alpha_0$ increases. With fixed $\alpha_0$, $r_h$ increases as $n$ increases, but $r_h$ decreases as $x$ increases.

Next we examine the motion of photons near these new regular BHs. The motion of these photons is governed by the Euler-Lagrange equation, which is expressed as follows
\begin{equation}
\label{eq:5}
\frac{d}{d \lambda}\left(\frac{\partial \mathscr{L}}{\partial \dot{x}^{\mu}}\right) = \frac{\partial \mathscr{L}}{\partial x^{\mu}}.
\end{equation}
Here, $\lambda$ is an affine parameter, and $\dot{x}^{\mu}$ represents the four-velocity of the photon. In the context of a static spacetime, the Lagrangian density $\mathscr{L}$ is expressed as
\begin{equation}
\label{eq:6}
\begin{split}
\mathscr{L} &= -\frac{1}{2} g_{\mu \nu} \frac{d x^{\mu}}{d \lambda} \frac{d x^{\nu}}{d \lambda} \\
&= -\frac{1}{2}\left[f(r) \dot{t}^{2} + f(r)^{-1}(r) \dot{r}^{2} + r^{2}\left(\dot{\theta}^{2} + \sin ^{2} \theta \dot{\phi}^{2}\right)\right].
\end{split}
\end{equation}

For photons, the Lagrangian density $\mathscr{L} = 0$. In a spherically symmetric spacetime, we analyze the motion of photons restricted to the equatorial plane of these new regular BHs by setting $\theta =\pi/2$ and $\dot\theta = 0$. Since the Lagrangian density is zero and the metric coefficients cannot be explicitly solved in terms of the $t$ and $\theta$ coordinates, there exist two conserved quantities, $E$ and $L$, representing the energy and angular momentum of the photons respectively
\begin{equation}
\label{eq:7}
    E = f(r)\left(\frac{dt}{d\lambda}\right),~~~\quad L = r^{2}\left(\frac{d\phi}{d\lambda}\right).
\end{equation}
By redefining the affine parameter as $\tilde{\lambda}\to\lambda/|L|$,the four-velocity can be expressed as
\begin{equation}
\label{eq:8}
\begin{gathered}
\frac{dt}{d\tilde{\lambda}} = \frac{1}{b f(r)}, ~~~
\frac{d\phi}{d\tilde{\lambda}} = \pm \frac{1}{r^{2}}, ~~~
\frac{dr}{d\tilde{\lambda}} = \sqrt{\frac{1}{b^{2}} - V_{\text{eff}}(r)}.
\end{gathered}
\end{equation}

Here, the $\pm$ signs denote the direction of photon motion on the equatorial plane: the ``$+$"  sign corresponds to clockwise motion, and the ``$-$" sign corresponds to counterclockwise motion. The impact parameter $b$ is defined as
\begin{equation}
\label{eq:9}
b = \frac{|L|}{E}.
\end{equation}
The effective potential $V_{\text{eff}}(r)$ is expressed as
\begin{equation}
\label{eq:10}
 \quad V_{\text{eff}}(r) = \frac{f(r)}{r^{2}}.
\end{equation}
For null geodesics, where $g_{\mu \nu} \dot{x}^{\mu} \dot{x}^{\nu} = 0$, we can obtain the first-order differential equation of motion
\begin{equation}
\label{eq:11}
\dot{r}^{2} + V_{\text{eff}} = \frac{1}{b^{2}}.
\end{equation}
Obviously, for photons to form a photon sphere, their circular orbit must satisfy $\dot{r}=0$ and $\ddot{r}=0$. Mathematically, this means
\begin{equation}
\label{eq:12}
V_{\text{eff}}^{\prime}\left(r_{c}\right) = 0, \quad V_{\text{eff}}\left(r_{c}\right) = \frac{1}{b_{c}^{2}}.
\end{equation}
Here, $r_{c}$ represents the radius of the photon sphere, and $b_{c}$ denotes the corresponding critical impact parameter, which is related to the shadow radius of these new regular BHs.

\end{multicols}
\begin{multicols}{2}
\begin{table}[H]
\setlength{\tabcolsep}{0.72em}
\caption{The locations of the outer horizon ($r_h$), photon sphere ($r_c$), and impact parameter ($b_c$) with the variation of the deviation parameter $\alpha_0$ ($M = 1$).}
\begin{tabular}{ccccc}
\hline\hline
$\alpha_0$ & Type & $r_h$ & $r_c$ & $b_c$ \\
\hline
\multirow{1}{*}{0} & Sch & 2.00000 & 3.00000 & 5.19615 \\
\hline
\multirow{2}{*}{0.3} & $x = 2/3, n = 2$ & 1.82833 & 2.81574 & 5.00833 \\
 & Bardeen & 1.83402 & 2.81929 & 5.01087 \\
\hline
\multirow{2}{*}{0.72} & $x = 2/3, n = 2$ & 1.33657 & 2.44678 & 4.66393 \\
 & Bardeen & 1.49244 & 2.48965 & 4.69073 \\
\hline\hline
\end{tabular}
\label{tab:1}
\end{table}
We present in Table \ref{tab:1} and \ref{tab:2} the numerical results for the event horizon $r_h$, the photon sphere $r_c$, and the critical impact parameter $b_c$ corresponding to various values of the parameter $\alpha_0$. The results indicate that the values of $r_h$, $r_c$, and $b_c$ of Schwarzschild BH are greater than those of these new regular BHs.  For a given $\alpha_0$, the values of $r_h$, $r_c$, and $b_c$ of these new regular BHs are related to the spacetime deformation parameters ($x$ and $n$) and the type of core they possess. As the parameter $\alpha_0$ increases, the values of $r_h$, $r_c$, and $b_c$ of these new regular BHs decrease. For a given $\alpha_0$, the values of $r_h$, $r_c$, and $b_c$ of these new regular BHs are related to the spacetime deformation parameters ($x$ and $n$) and the type of core they possess. When the core type is the same, these values increase with the increase of the dimensionless parameters ($x$ and $n$). This means that for the new regular BH, when $x=1$ and $n=3$, their values are greater than those when $x=2/3$ and $n=2$,and the values of Hayward BH are greater than those of Bardeen BH. When the dimensionless parameters ($x$ and $n$) are the same, the traditional regular BHs with a dS core are slightly greater values than these new regular BHs with a Minkowski core. This discrepancy becomes more significant as the parameter $\alpha_0$ increases. This means that Bardeen BH with a dS core is slightly greater values than the one with a Minkowski core ($x=2/3$ and $n=2$). similarly, Hayward BH with a dS core is slightly greater values than the one with a Minkowski core ($x=2/3$ and $n=2$).

\begin{table}[H]
\setlength{\tabcolsep}{0.8em}
\caption{The locations of the event horizon, the photon sphere, and the critical impact parameter of the new regular BH ($x = 1, n = 3$) and the Hayward BH with the various $\alpha_0$. We set $M=1$.}
\begin{tabular}{ccccc}
\hline\hline
$\alpha_0$ & Type & $r_h$ & $r_c$ & $b_c$ \\
\hline
\multirow{2}{*}{0.3} & $x = 1, n = 3$ & 1.91656 & 2.92900 & 5.13562 \\
 & Hayward & 1.91849 & 2.92968 & 5.13600 \\
\hline
\multirow{2}{*}{0.72} & $x = 1, n = 3$ & 1.74762 & 2.80996 & 5.03894 \\
 & Hayward & 1.77024 & 2.81540 & 5.04179 \\
\hline
\multirow{2}{*}{0.92} & $x = 1, n = 3$ & 1.59272 & 2.74067 & 4.98595 \\
 & Hayward & 1.67020 & 2.75161 & 4.99145 \\
\hline\hline
\end{tabular}
\label{tab:2}
\end{table}

\begin{figure}[H]
\centering
\subfloat[ \(x=2/3\), \(n=2\)]{
    \includegraphics[width=0.22\linewidth]{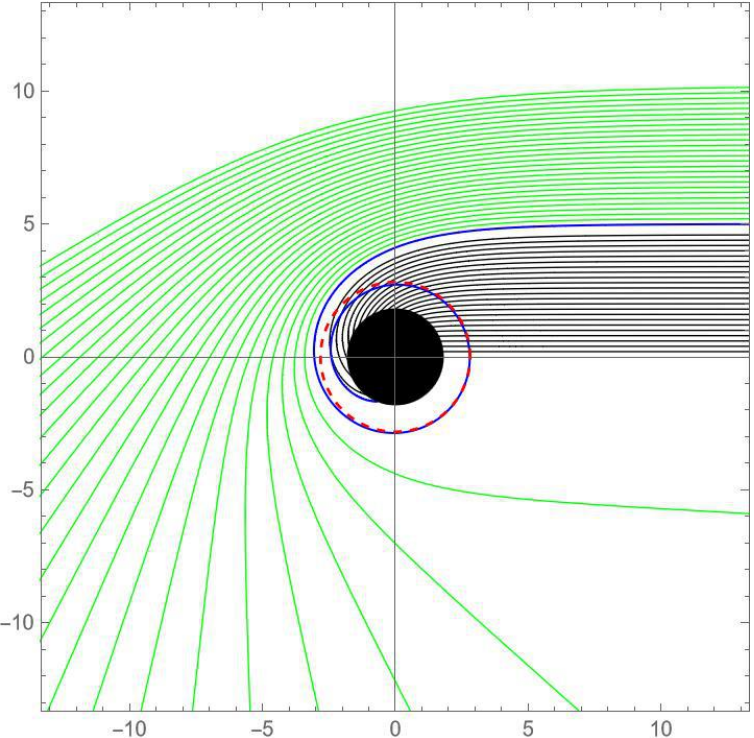}
    \label{fig:2a}
}
\hspace{2pt} 
\subfloat[ Bardeen BH]{
    \includegraphics[width=0.22\linewidth]{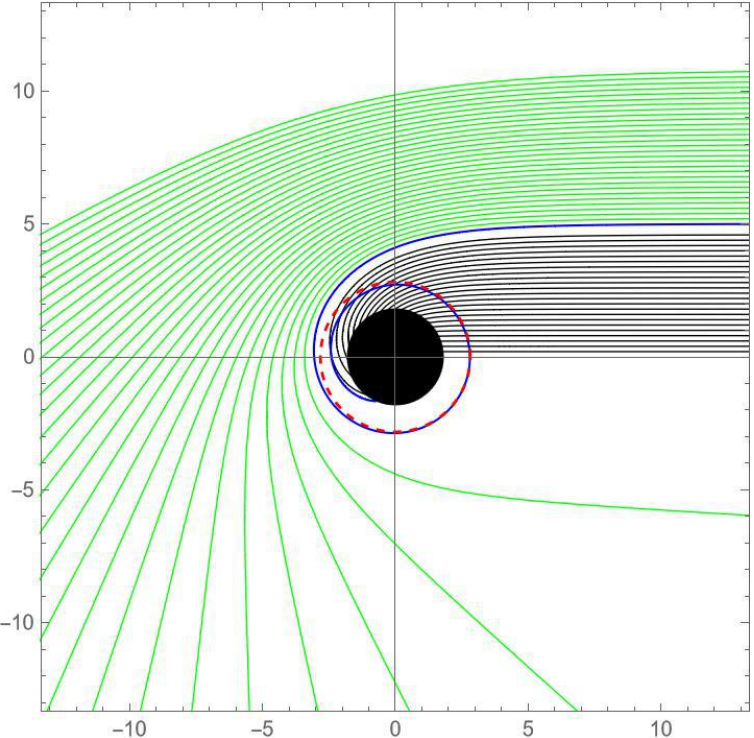}
    \label{fig:2b}
}
\hspace{2pt}
\subfloat[ \(x=1\), \(n=3\)]{
    \includegraphics[width=0.22\linewidth]{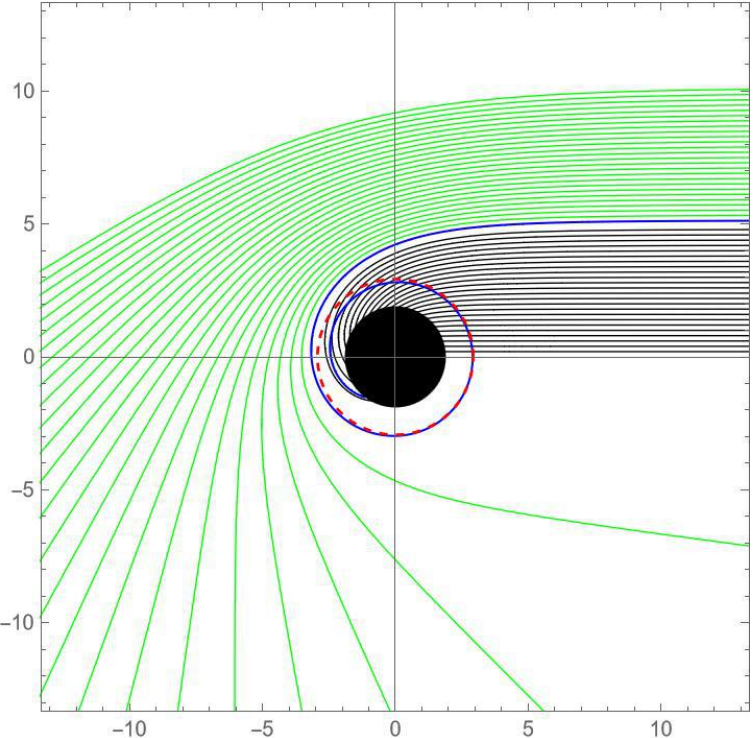}
    \label{fig:2c}
}
\hspace{2pt}
\subfloat[ Hayward BH]{
    \includegraphics[width=0.22\linewidth]{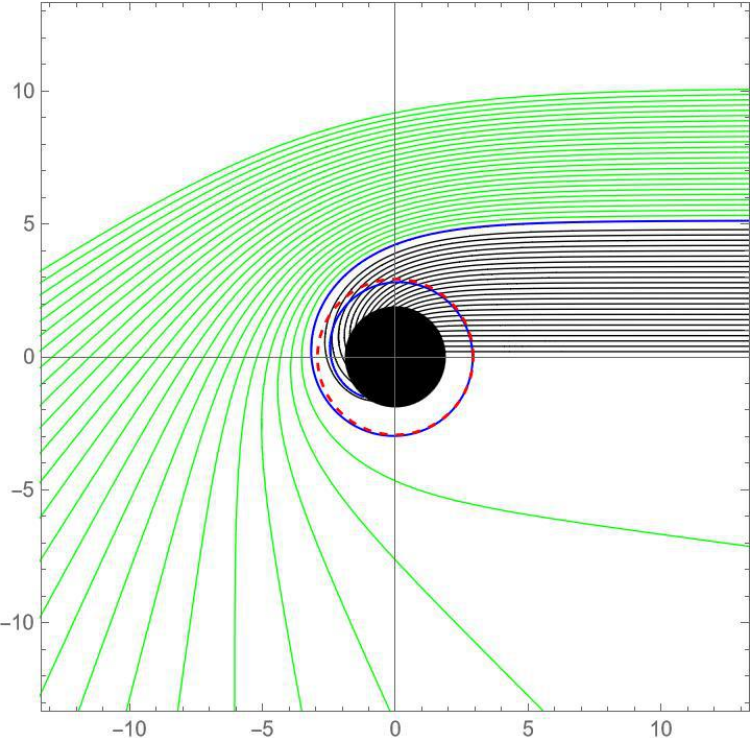}
    \label{fig:2d}
}
\\ 
\subfloat[ \(x=2/3\), \(n=2\)]{
    \includegraphics[width=0.22\linewidth]{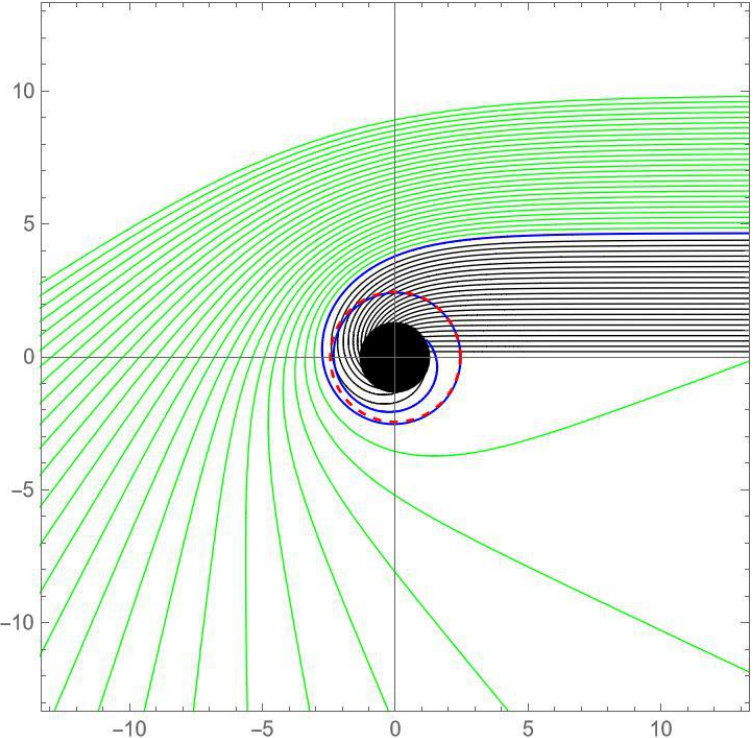}
    \label{fig:2e}
}
\hspace{2pt}
\subfloat[ Bardeen BH]{
    \includegraphics[width=0.22\linewidth]{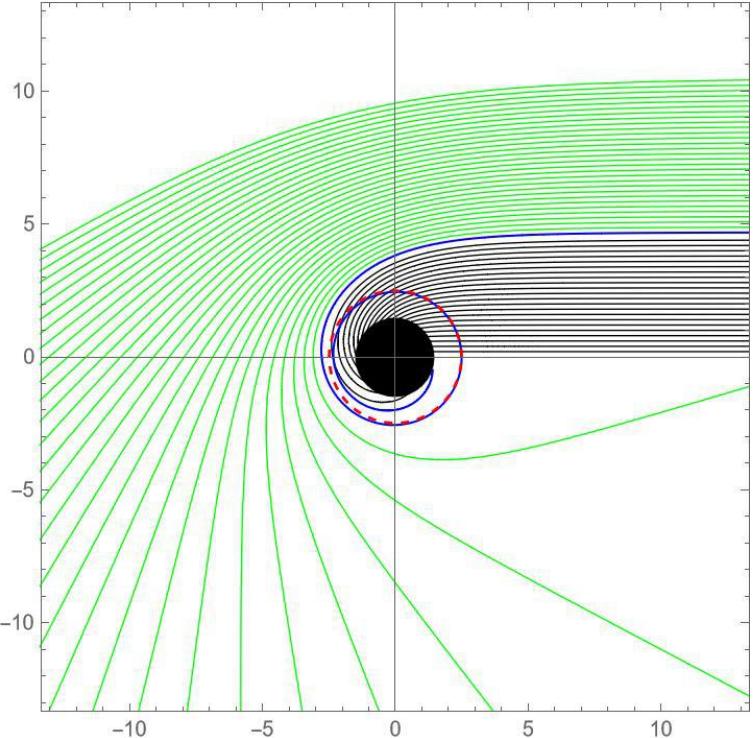}
    \label{fig:2f}
}
\hspace{2pt}
\subfloat[ \(x=1\), \(n=3\)]{
    \includegraphics[width=0.22\linewidth]{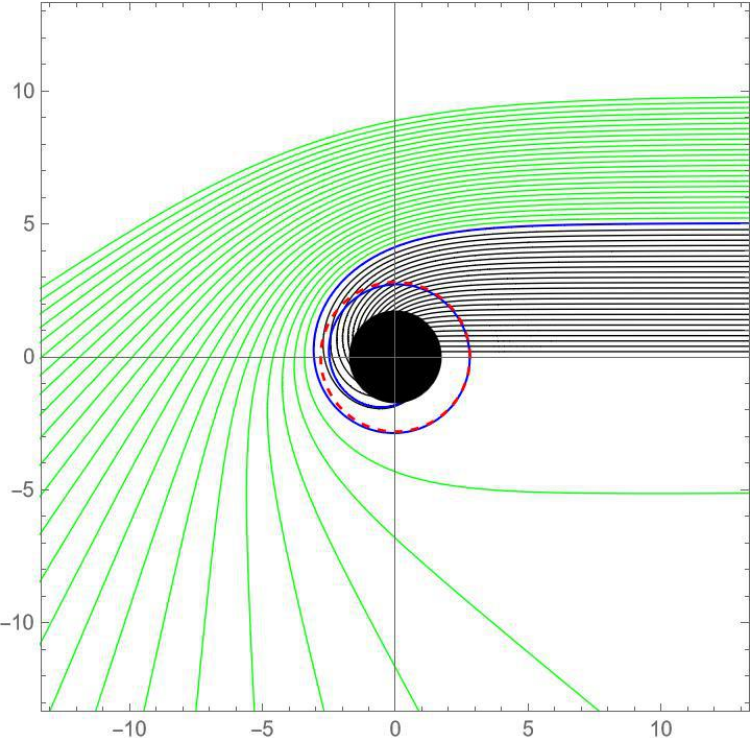}
    \label{fig:2g}
}
\hspace{2pt}
\subfloat[Hayward BH]{
    \includegraphics[width=0.22 \linewidth]{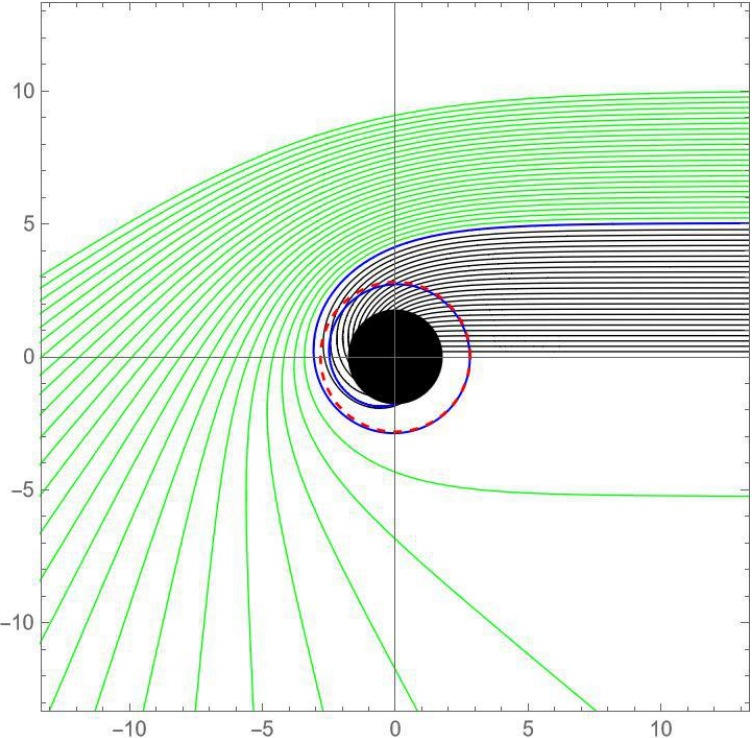}
    \label{fig:2h}
}
\caption{ The trajectories of light rays for the regular BHs with different cores. The top and bottom plots correspond to $\alpha_0=0.3$ and $\alpha_0=0.72$, respectively. We set $M=1$.}
\label{fig:2}
\end{figure}

For better understanding of how the specific paramet- ers affect the photon trajectories and their classification, we describe the motion of photons around these new reg- ular BHs by tracking the change in the radial coordinate as a function of the azimuthal angle $\phi$. By introducing the parameter $u = 1/r$, we are able to derive the trajectory equation
\begin{equation}
\label{eq:13}
\frac{d u}{d \phi}=\sqrt{\frac{1}{b^{2}}-u^{2}\left(1-2 M u e^{-\alpha_{0} M u^{n}}\right)}.
\end{equation}
and the azimuthal angle $\phi$ can be integrated out as
\begin{equation}
\label{eq:14}
\phi = \int \frac{1}{\sqrt{b^{-2}  - u^{2} \left(1 - 2 M u e^{-\alpha_{0} Mu^{n} } \right)}} du
\end{equation}
Based on the above formulas, we numerically simulated the turning points of photon trajectories around these new regular BHs and analyzed the corresponding geodesic geometry. To accurately distinguish among different light trajectories, we used polar coordinates $(b, \phi)$ and employed a ray tracing code to plot the paths of light as it moves around these new regular BHs, as shown in Fig.\ref{fig:2}. These plots demonstrate the changes in trajectories for different impact parameters $b$, with the distance between the different impact parameters $b$ being $0.2$.

In Fig.\ref{fig:2}, the black disk represents the area inside the event horizon, and the red dashed circle denotes the trajectory of light rays with the critical parameter $b=b_c$ which finally forms the photon sphere. For $b < b_c$, the light trajectories are engulfed by these regular BHs, making them invisible to an observer at infinity, as shown by the black lines in Fig.\ref{fig:2}. For $b > b_c$, the light trajectories are deflected, enabling them to reach an observer at infinity, as shown by the green lines in Fig.\ref{fig:2}. Moreover, as the parameter $\alpha_0$ increases, the area in which the light trajectories are captured by these regular BHs decreases, as shown in Table \ref{tab:1} and \ref{tab:2}. Additionally, the area in which the light trajectories are captured by these regular BHs is also related to the dimensionless parameters ($x$ and $n$) and the type of core they possess, as shown in Table \ref{tab:1} and \ref{tab:2}. Fig.\ref{fig:2} shows that for the same core, the shadow radius of the new regular BH with the parameters $x=1$ and $n=3$ is greater than that with $x=2/3$ and $n=2$, and Hayward BH has a greater shadow radius than Bardeen BH. For the same dimensionless parameters ($x$ and $n$), the shadow radius of Bardeen BH with a dS core is slightly greater than that with a Minkowski core ($x=2/3$ and $n=2$), and the shadow radius of Hayward BH with a dS core is slightly greater than that with a Minkowski core ($x=2/3$ and $n=2$).

These results indicate that the changes in the paramet- ers induced by quantum gravity effect (represented by $\alpha_0$) and spacetime deformation (characterized by $x$ and $n$) can modify the radius of these shadows. In particular, as the parameter $\alpha_0$ increases, the differences in the motion of photons around these new regular BHs with different parameters ($x$ and $n$) and distinct cores become more evident. This finding offers crucial clues for further research on the impact of quantum gravity and spacetime structure on the images of the new regular BHs presented in the next section under spherical accretion models.

\end{multicols}
\begin{multicols}{2}

\section{Images of the new regular BHs with a Minkowski core} \label{sec3}
In this section, we investigate the shadow images of these new regular BHs with a Minkowski core, considering two models of spherically symmetric accretion: static spherical accretion and infalling spherical accretion. We will examine the impact of the parameter ($\alpha_0$) that is induced by quantum gravity effects and the parameter ($n$) that describes the formation of various spacetime structures on the observational characteristics of the shadow.

\subsection{Static spherical accretion} \label{sec3-1}

Generally, when matter in the universe is captured by a BH, the morphology of the accretion flow around the BH is governed by its angular momentum. High angular momentum causes the matter to form a disk-like accretion, whereas with extremely low angular momentum, the matter flows radially towards the BH, leading to a spherically symmetric accretion \cite{80,81,82,83,84,85,86}. We first investigate the shadow images and photon spheres of these new regular BHs with a Minkowski core surrounded by the static spherical accretion. This assumes that the accretion is static relative to these new regular BHs. For an observer at infinity, the observed specific intensity $I_{obs}(\nu_o)$ (measured in $\text{erg}$ $\text{s}^{-1}$ $\text{cm}^{-2}$ $\text{str}^{-1}$ $\text{Hz}^{-1}$) is determined by integrating the specific emissivity along the photon path $\gamma$ \cite{85,86}, i.e.
\begin{equation}
\label{eq:15}
I_{obs}\left(\nu_{o}\right)=\int_{\gamma} g^{3} j_{e}\left(\nu_{e}\right) d l_{p r o p}.
\end{equation}
Here, $\nu_{o}$ is the frequency of the observed photon, and $\nu_{e}$ is the frequency of the emitted photon. The redshift factor $g$ is defined as $g = \nu_{o} / \nu_{e} = f(r)^{1/2}$. In the rest frame of the emitter, $j_{e}(\nu_{e})$ represents the emissivity per unit volume and is typically given by the form $j_{e}(\nu_{e}) \propto \delta(\nu_{r} - \nu_{e}) / r^{2}$, where $\nu_{r}$ is the frequency in the emitter's rest frame \cite{86}. The infinitesimal proper length $dl_{\text{prop}}$ is given by
\begin{equation}
\label{eq:16}
d l_{\text {prop }} = \sqrt{\frac{1}{f(r)} d r^{2}+r^{2} d \phi^{2}} = \sqrt{\frac{1}{f(r)}+r^{2}\left(\frac{d \phi}{d r}\right)^{2}} d r.
\end{equation}
where $d \phi /d r$ is obtained by Eq.(\ref{eq:13}). Subsequently, the observed specific intensity is expressed as
\begin{equation}
\label{eq:17}
I_{obs} = \int_{\gamma} \frac{f(r)^{3/2}}{r^{2}}\sqrt{\frac{1}{f(r)}+r^{2}\left(\frac{d \phi}{d r}\right)^{2}} d r.
\end{equation}

\begin{figure}[H] 
    \centering
    \includegraphics[width=0.45\textwidth]{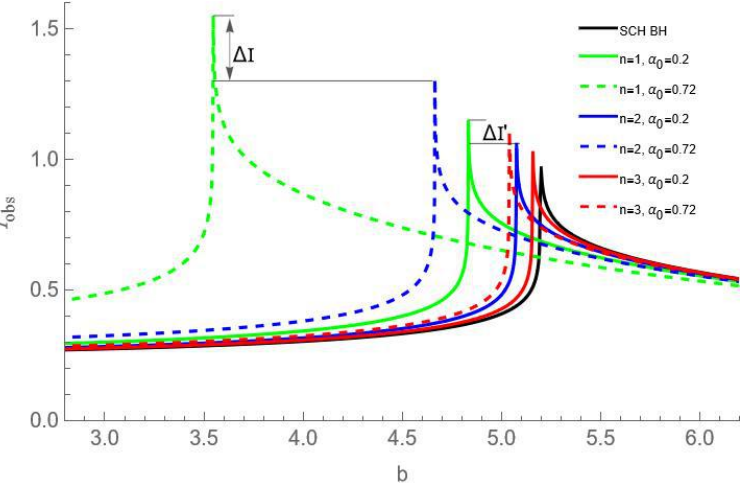} 
    \caption{ The variation of the observed specific intensity radiated from the static spherical accretion with the impact parameter under different values of $\alpha_0$ and $n$. We set $M = 1$.} 
    \label{fig:3} 
\end{figure}

According to Eq.(\ref{eq:17}), the observed specific intensity $I_{obs}$ as a function of impact parameter $b$. It is noteworthy that we aim to study how variations in the quantum gravity effect parameter ($\alpha_0$) and spacetime deformation ($n$) affect the observed specific intensity. We compared the cases with $x=2/3$ and $n=2$ to those with $x=1$ and $n=3$. In Fig.\ref{fig:3}, we illustrate the trend of the observed specific intensity at spatial infinity as a function of the impact parameter $b$. It is evident that the observed specific intensity increases gradually with $b$ when $b<b_c$, but it rises sharply as it approaches $b_c$, reaching a maximum at this position. Following this, as the value of $b$ continues to increase beyond $b_c$, the observed specific intensity begins to decrease gradually, eventually approaching zero at infinity.

Fig.\ref{fig:3} shows that the peak intensity for these new regular BHs with a Minkowski core is consistently higher than that of Schwarzschild BH. When the spacetime deformation $n=1$, the peak intensity for the new regular BH corresponding to $\alpha_0 = 0.72$ (shown as the green dashed line) is greater than the case with $\alpha_0 = 0.2$ (shown as the green solid line). Similarly, for the spacetime deformation factors $n=2$ and $n=3$, the peak intensity corresponding to $\alpha_0 = 0.72$ (shown as the blue and red dashed lines) is also greater than those for $\alpha_0 = 0.2$ (shown as the blue and red solid lines). These results imply that the larger the value of $\alpha_0$, the stronger the observed specific intensity of these new regular BHs. Notably, the parameter $\alpha_0$ arises from the corrections due to quantum gravity effect, indicating that the stronger the quantum gravity effect, the higher the luminosity of these new regular BHs.

Furthermore, Fig.\ref{fig:3} illustrates that for the parameter $\alpha_0=0.72$, the peak intensity decreases in the order of $n = 1$, $n=2$, and $n=3$, corresponding to the green dashed line, blue dashed line, and red dashed line, respectively. Similarly, for the parameter $\alpha_0=0.2$, the peak intensity also decreases sequentially for $n=1$, $n=2$, and $n=3$, corresponding to the green solid line, blue solid line, and red solid line, respectively. This implies that as the parameter $n$ increases, the observed specific intensity of these new regular BHs gradually weakens, indicating that a smaller value of $n$ can increase the optical brightness and enable the observer to detect more luminous signals.

\begin{figure}[H]
   \centering
   \subfloat[$n=1$, $\alpha_0=0.2$]{%
       \includegraphics[width=0.22\textwidth]{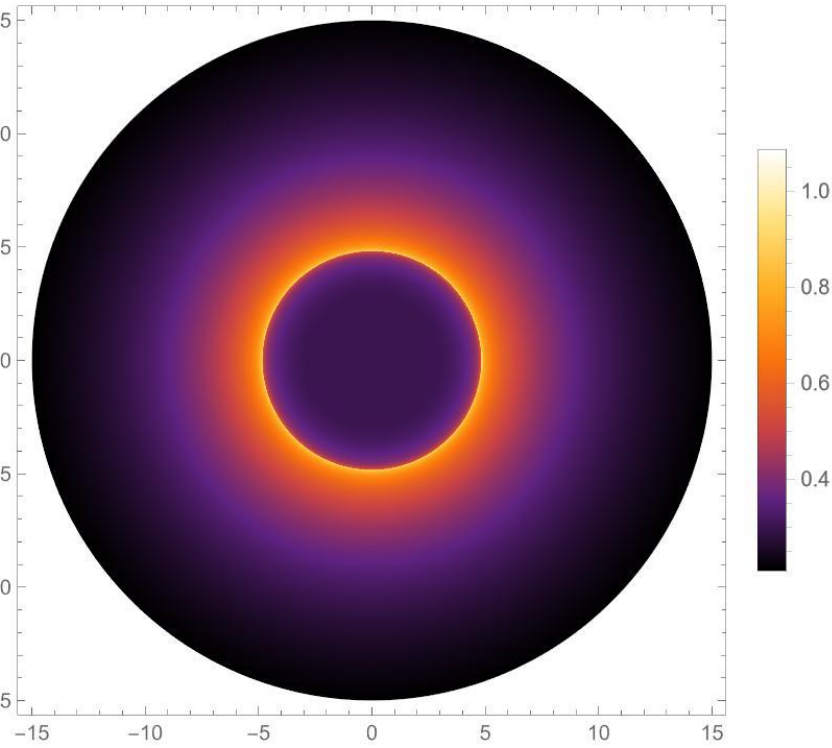}
       \label{fig:4a}
   }
   \hspace{0.25cm}
   \subfloat[$n=2$, $\alpha_0=0.2$]{%
       \includegraphics[width=0.22\textwidth]{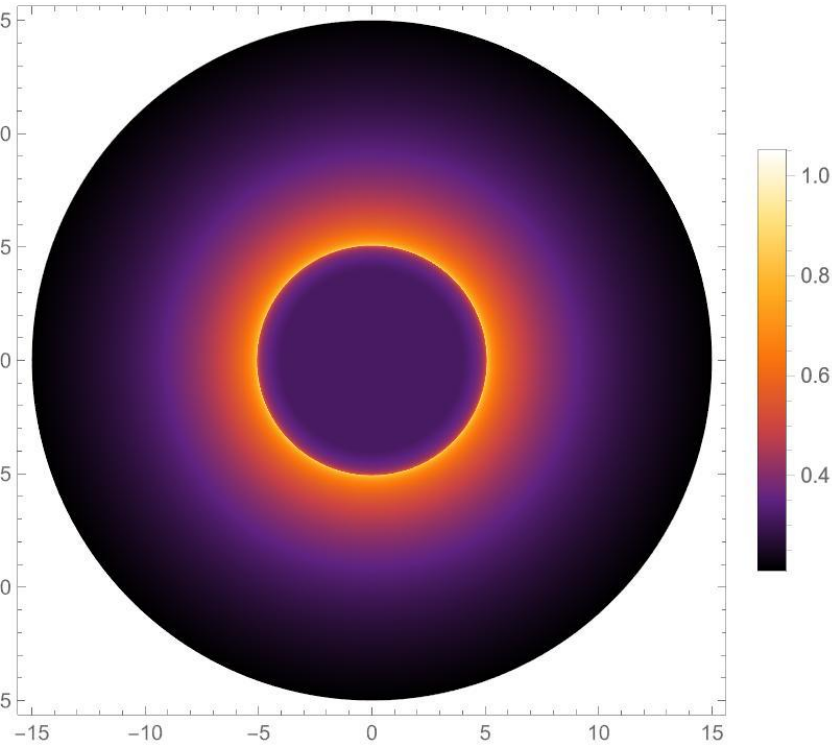}
       \label{fig:4b}
   }
   \hspace{0.25cm}
   \subfloat[$n=1$, $\alpha_0=0.72$]{%
       \includegraphics[width=0.22\textwidth]{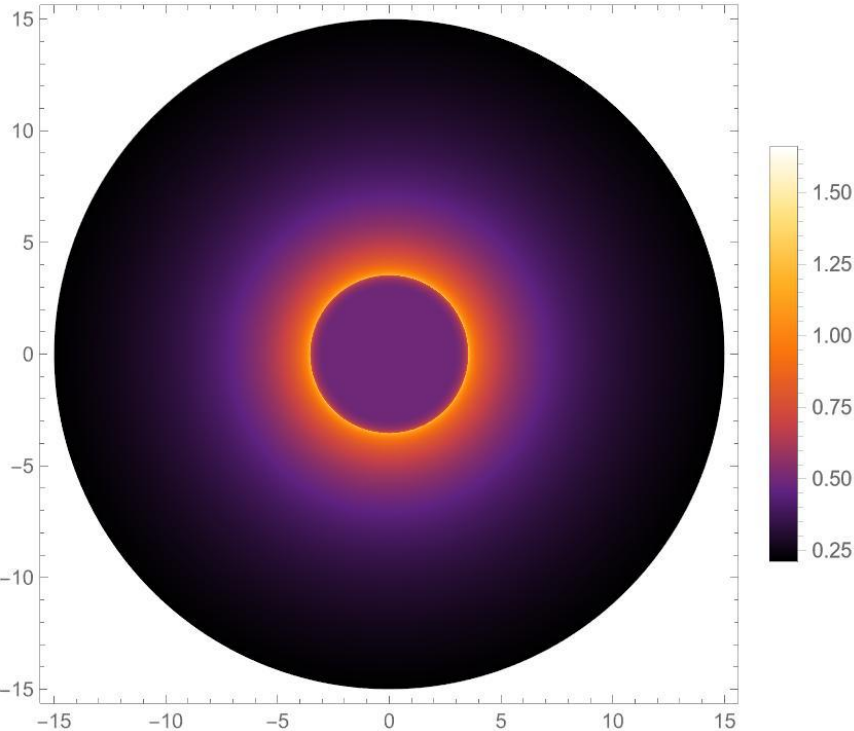}
       \label{fig:4c}
   }
   \hspace{0.25cm}
   \subfloat[$n=2$, $\alpha_0=0.72$]{%
       \includegraphics[width=0.22\textwidth]{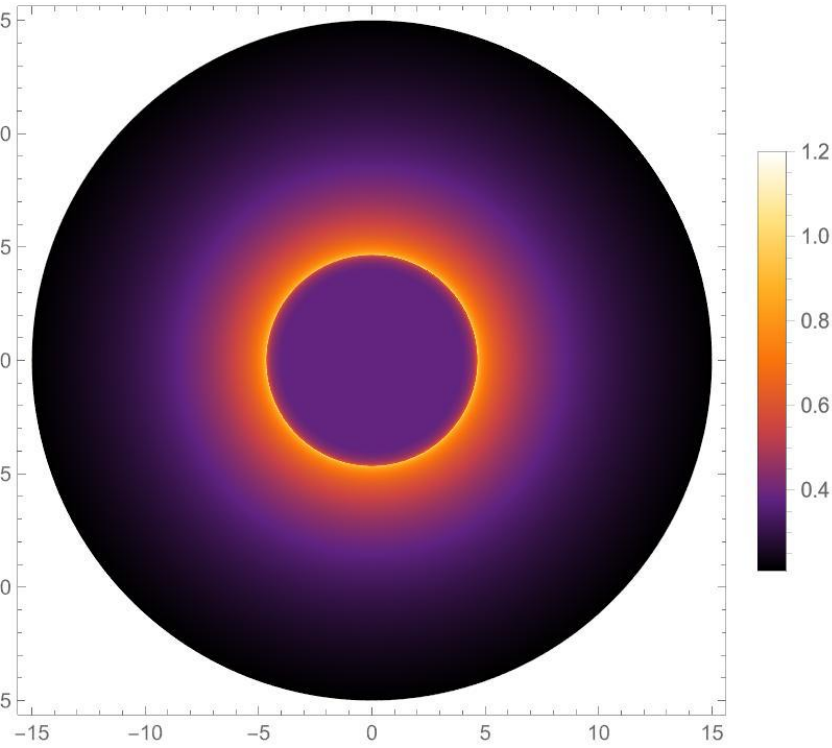}
       \label{fig:4d}
   }
   \caption{The optical appearances of these new regular BHs under the static spherical accretion for different values of $\alpha_0$ and $n$.}
   \label{fig:4}
\end{figure}

Finally, in Fig.\ref{fig:3}, $\Delta I$ represents the difference in the peak intensities for $n=1$ and $n=2$ corresponding to $\alpha_0 = 0.72$, and $\Delta I'$ represents the difference $n=1$ and $n=2$ corresponding to $\alpha_0 = 0.2$. It is evident that $\Delta I$ is greater than $\Delta I'$, which means that the difference between the peak intensities becomes more pronounced as the parameter $\alpha_0$ increases. This indicates that the larger $\alpha_0$, the easier it is to distinguish between these new regular BHs with different $n$ through the observed specific intensity. We take $n = 1$ and $n = 2$ as examples and present the two-dimensional image of the observed specific intensity for these new regular BHs with different $\alpha_0$ in Fig.\ref{fig:4}.

In Fig.\ref{fig:4}, the photon sphere is at the bright ring with the strongest luminosity. It can be observed that the inner region of the photon sphere is not completely dark, as a small fraction of the radiation can escape from these new regular BHs, so a weak luminosity can be observed near the photon sphere. Obviously, the parameters $\alpha_0$ and $n$ have a significant impact both on the photon ring brightness and the shadow radius. In Fig.\ref{fig:4}, we can see that for a given $n$, the larger $\alpha_0$ leads to a brighter photon ring of the BH image, but a smaller radius of the BH shadow and photon sphere. For a given $\alpha_0$, the larger $n$ leads to a dimmer photon ring of the BH image, but a larger radius of the BH shadow and the photon sphere.

\end{multicols}
\begin{multicols}{2}
In addition, as $\alpha_0$ increases, the differences between the shadow images for $n=1$ and $n=2$ become more apparent in Fig.\ref{fig:4}. Therefore, under the static spherical accretion model, the enhancement of quantum gravity effect can increase the luminosity of photon ring and enlarge the shadow radius for these new regular BHs. As $n$ decreases, the enhancement of quantum gravity effect becomes more pronounced. This is because a smaller $n$ amplifies the influence of $\alpha_0$ on the optical characteristics of these new regular BHs, thereby making it easier for the observer to distinguish these new regular BHs.

\subsection{Infalling spherical accretion} \label{sec3-2}

In this section, we will investigate the shadow images of these new regular BHs with a Minkowski core surrounded by an infalling spherical accretion. This is a dynamical model where the accretion is assumed to move radially towards the BH. In this case, the observed specific intensity in Eq.(\ref{eq:15}) is still applicable, but the corresponding redshift factor differs from the one in static spherical accretion. The redshift factor in the infalling spherical accretion can be expressed as
\begin{equation}
\label{eq:18}
g_{i} = \frac{{K}_{\rho} u_{0}^{\rho}}{{K}_{\sigma} u_{e}^{\sigma}}, \quad {K}^{\mu} = \dot{x}_{\mu}.
\end{equation}
Here,$ \quad {K}^{\mu}$ represents the four-momentum of a photon emitted by the accreting matter. $u_{0}^{\rho} = (1, 0, 0, 0)$ corresponds to the four-velocity of a static observer, and $u_{e}^{\sigma}$ denotes the four-velocity of the infalling accretion, given by
\begin{equation}
\label{eq:19}
u_{e}^{t}=\frac{1}{f(r)}, \quad u_{e}^{r}=-\sqrt{1-f(r)}, \quad u_{e}^{\theta}=u_{e}^{\phi}=0.
\end{equation}
The four-momentum of the photon is derived from the null geodesic, which is
\begin{equation}
\label{eq:20}
K_{t}=\frac{1}{b}, \quad K_{r}= \pm \frac{1}{f(r)} \sqrt{\frac{1}{b^{2}}-\frac{f(r)}{r^{2}}}.
\end{equation}
It is noted that in $K_{r}$, the $+$ or $-$ sign indicates whether the photon is moving radially inward or outward from the BH. For the case of the infalling spherical accretion, the redshift factor can be written as
\begin{equation}
\label{eq:21}
g_i = \left[ u_e^t + \left( \frac{{K}_r}{{K}_t} \right) u_e^r \right]^{-1}.
\end{equation}
Furthermore, unlike the static spherical accretion case, the proper distance $dl_{\text{prop}}$ is defined as the spatial distance measured in the local rest frame of the infalling spherical accretion. The four-momentum $K^\mu=dx^\mu/d\lambda$ (with $K^\mu K_\mu=0$) is projected onto the spatial hypersurface orthogonal to $u_e^\mu$ using the projection tensor $h_\nu^\mu=\delta_\nu^\mu + u_e^\mu u_{e\nu}$, so we have
\begin{equation}
\label{eq:21-1}
K_{\perp}^\mu=h_\nu^\mu K^\nu = K^\mu + u_e^\mu(u_{e\nu} K^\nu).
\end{equation}
In the local rest frame, the magnitude of the spatial momentum $\vert K_{\perp}^\mu \vert$ is given by
\begin{equation}
\label{eq:21-2}
\vert K_{\perp}^\mu \vert=\sqrt{g_{\mu\nu} (K_{\perp}^\mu) (K_{\perp}^\nu)}=\sqrt{(K_\mu u_e^\mu)^2}=\vert K_\mu u_e^\mu \vert.
\end{equation}
Since $K^\mu$ and $u_e^\mu$ are null and infalling accretion respectively, their inner product $K_\mu u_e^\mu < 0$. Hence, $\vert K_{\perp}^\mu \vert=-K_\mu u_e^\mu$. The proper distance $dl_{\text{prop}}$ traveled by the photon in the local rest frame over an affine parameter interval $d\lambda$ is
\begin{equation}
\label{eq:21-3}
dl_{\text{prop}}=\vert K_{\perp}^\mu \vert d\lambda=-K_\mu u_e^\mu d\lambda.
\end{equation}
By convention, the negative sign is absorbed into the definition, yielding
\begin{equation}
\label{eq:22}
dl_{\text{prop}} =K_{\mu} u_{e}^{\mu} d \lambda=\frac{K_{t}}{g_{i}\left|K_{r}\right|} d r.
\end{equation}
The observed specific intensity for the infalling spherical accretion is calculated as follows
\begin{equation}
\label{eq:23}
I_{\text{obs}}=\int_{\gamma} \frac{g_{i}^{3} K_{t}}{r^{2}\left|K_{r}\right|} d r.
\end{equation}
Based on the above equation, we obtain the variation of the observed specific intensity $I_{\text{obs}}$ with the impact parameter $b$ under different parameters $\alpha_0$ and $n$, observed by an observer at infinity.

\begin{figure}[H] 
    \centering
    \includegraphics[width=0.45\textwidth]{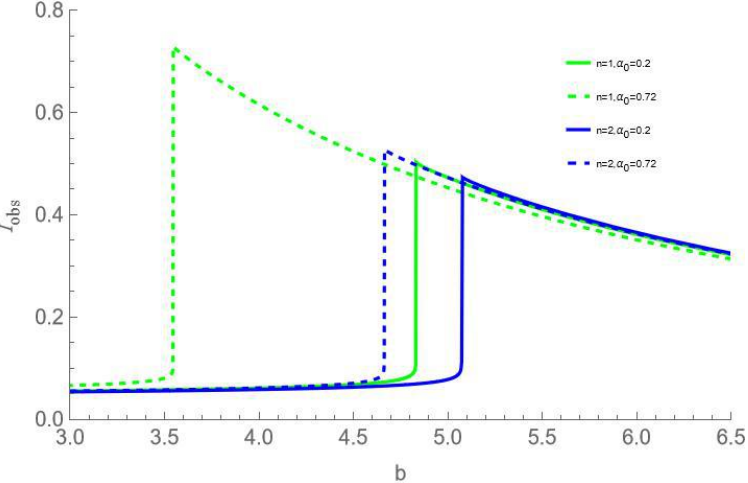} 
    \caption{  The variation of the observed specific intensity radiated from the infalling spherical accretion with the impact parameter under different values of $\alpha_0$ and $n$. We set $M = 1$.} 
    \label{fig:5} 
\end{figure}

We select the cases of $\alpha_0 = 0.2$ and $\alpha_0 = 0.72$, and $n = 1$ and $n = 2$ to draw Fig.\ref{fig:5}. It can be found that as the impact parameter $b$ increases, the observed specific intensity increases slowly. When the impact parameter $b$ approaches the critical value $b_c$, the observed specific intensity increases rapidly and reaches a peak. Subsequently, as $b$ continues to increase, the observed specific intensity gradually decreases. At the same time, the variation of the observed specific intensity is closely related to the parameters $n$ and $\alpha_0$. For a fixed $n$, with the increase of $\alpha_0$, the impact parameters $b$ corresponding to the peak position of the observed specific intensity decreases, and the peak intensity increases. For a fixed $\alpha_0$, with the increase of $n$, the impact parameters $b$ corresponding to the peak position of the observed specific intensity increases, and the peak intensity decreases.

\begin{figure}[H]
   \centering
   \captionsetup{justification=raggedright, singlelinecheck=false}
   \subfloat[$n=1$, $\alpha_0=0.2$]{%
       \includegraphics[width=0.22\textwidth]{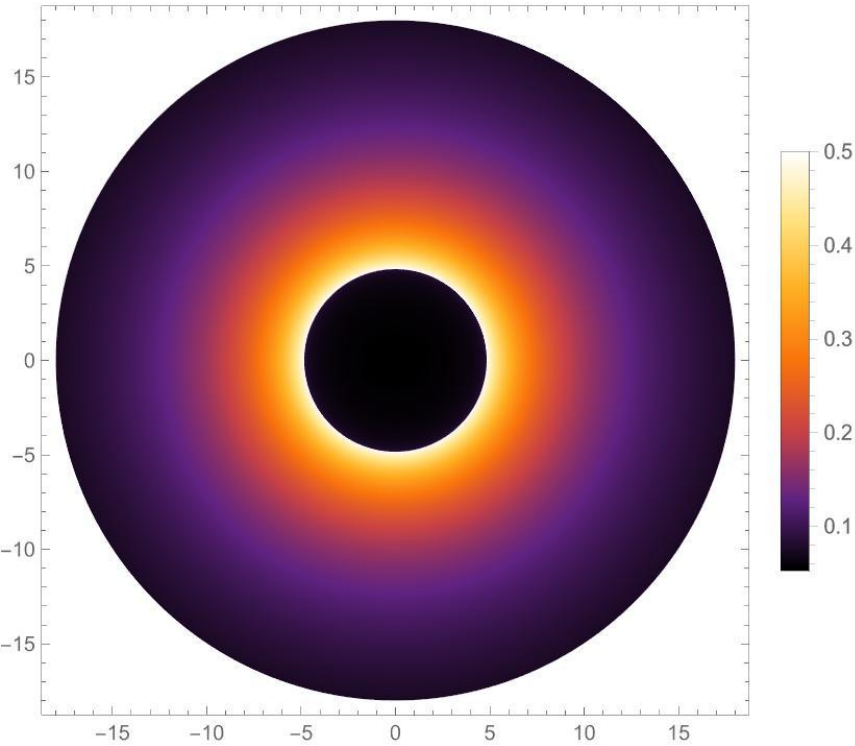}
       \label{fig:6a}
   }
   \hspace{0.25cm}
   \subfloat[$n=2$, $\alpha_0=0.2$]{%
       \includegraphics[width=0.22\textwidth]{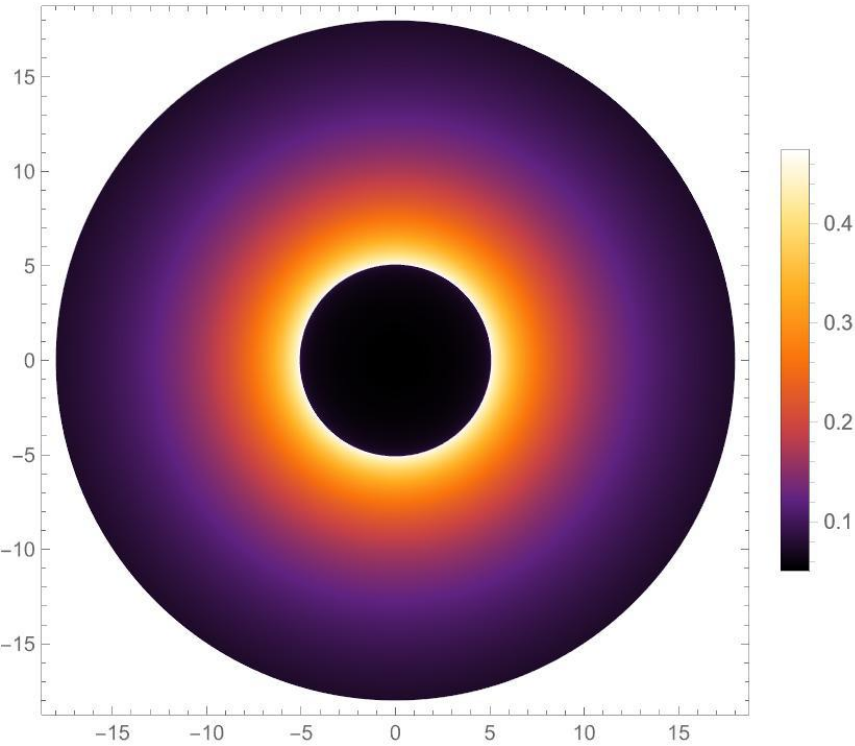}
       \label{fig:6b}
   }
   \hspace{0.25cm}
   \subfloat[$n=1$, $\alpha_0=0.72$]{%
       \includegraphics[width=0.22\textwidth]{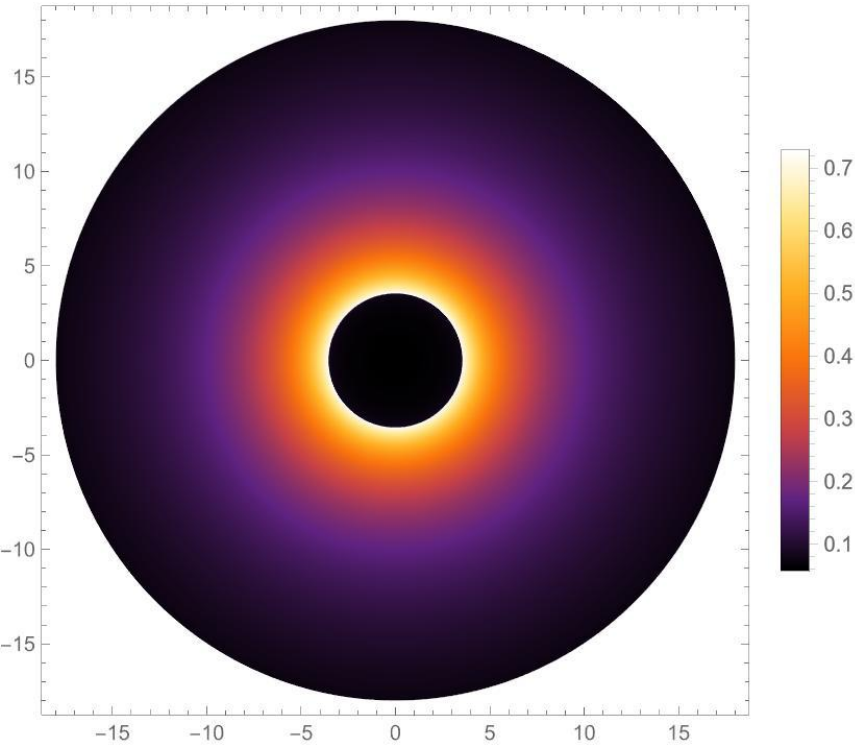}
       \label{fig:6c}
   }
   \hspace{0.25cm}
   \subfloat[$n=2$, $\alpha_0=0.72$]{%
       \includegraphics[width=0.22\textwidth]{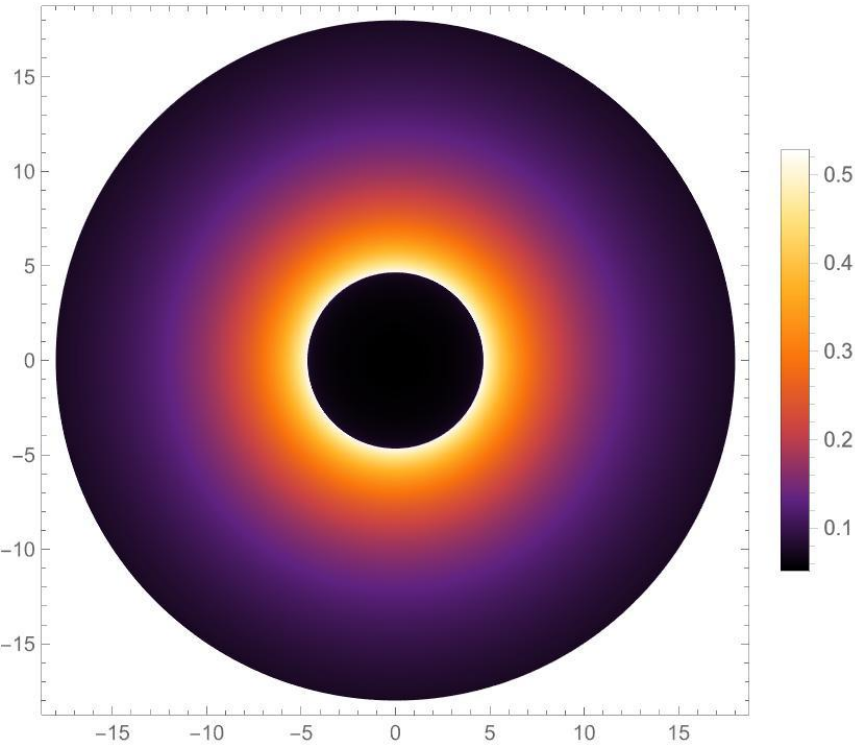}
       \label{fig:6d}
   }
   \caption{The optical appearances of these new regular BHs under the infalling spherical accretion for different values of $\alpha_0$ and $n$.}
   \label{fig:6}
\end{figure}

Fig.\ref{fig:6} shows the optical appearances of these new regular BHs under the infalling spherical accretion. By comparing Figs.(\ref{fig:6a}) and (\ref{fig:6c}), it can be found that as $\alpha_0$ increases, the peak intensity becomes higher, while the shadow radius and the photon sphere radius become smaller. Conversely, by comparing Figs.(\ref{fig:6c}) and (\ref{fig:6d}), as the parameter $n$ increases, the peak intensity becomes smaller, and the shadow radius and the photon sphere radius become larger. At the same time, by comparing the differences in the photon ring and the shadow radius between Figs.(\ref{fig:6a}) and (\ref{fig:6c}), as well as between Figs.(\ref{fig:6b}) and (\ref{fig:6d}), it can be found that as the parameter $\alpha_0$ increases, the differences in the optical appearances of $n=1$ and $n=2$ become more obvious. Therefore, under the infalling spherical accretion, the larger $\alpha_0$ is, the greater the luminosity of the photon ring, but the smaller the shadow radius is. Conversely, as the parameter $n$ increases, the photon ring becomes dimmer, and the shadow radius becomes larger.

Combining the results in Fig.\ref{fig:4} under the static spherical accretion and those in Fig.\ref{fig:6} under the infalling spherical accretion, for an observer at infinity, the observed specific intensity, the shadow radius, and the photon sphere radius of these new regular BHs with a Minkowski core are closely related to the intrinsic parameters of the BH themselves. Under both spherical accretion, the larger $\alpha_0$ is, the larger  observed specific intensity, while the shadow radius and the photon sphere radius decrease.  In contrast, the larger $n$ is, the smaller observed specific intensity, while the shadow radius and the photon sphere radius increase. Then, by comparing the results of the two spherical accretion, it is found that the shadow radius and the location of the photon sphere is independent of the choice of the spherical accretion model. However, the observed specific intensity of these new regular BHs is related to the choice of the spherical accretion model. The observed specific intensity obtained from the static spherical accretion is much larger than that from the infalling spherical accretion. This indicates that the BH shadow can represent the intrinsic properties of the BH spacetime, and the choice of the accretion model only affects the observed specific intensity of the shadow, not the structure of the BH shadow itself. Therefore, the spacetime characteristics of these new regular BHs can be distinguished by observing and studying the BH shadow.
\end{multicols}
\begin{multicols}{2}

In the next section, we will further compare the shadows formed by these new regular BHs with a Minkowski core and the traditional regular BHs with a dS core under the two different spherical accretion, and discuss the application of shadow research in distinguishing regular BHs with different cores.

\section{Comparison with the traditional regular BHs} \label{sec4}

As previously mentioned, these new regular BHs exhibit similar asymptotic behavior to the traditional Bardeen/Hayward BHs on large scales, but they differ in the core of the central region. Specifically, these new regular BHs have a Minkowski core, while the traditional Bardeen/Hayward BHs have a dS core. It has been found that in the thin disk accretion model of \cite{15}, the optical observational characteristics of the new regular BH and the corresponding traditional Bardeen BH may differ. Based on this, in this section, we will further compare the shadows and optical appearances of these new regular BHs with those of the corresponding traditional Bardeen/Hayward BHs under the spherical accretion model. Meanwhile, the research in the previous section shows that different spherical accretion models can affect the observed specific intensities of the shadows. Therefore, we will also compare the shadows of these new regular BHs and the corresponding traditional regular BHs under two different spherical accretion models.

\subsection{Comparison under static spherical accretion} \label{sec4-1}

\begin{figure}[H]
    \centering
    \subfloat[fixing $n=2$ and changing $\alpha_0$]{%
        \includegraphics[width=0.38\textwidth]{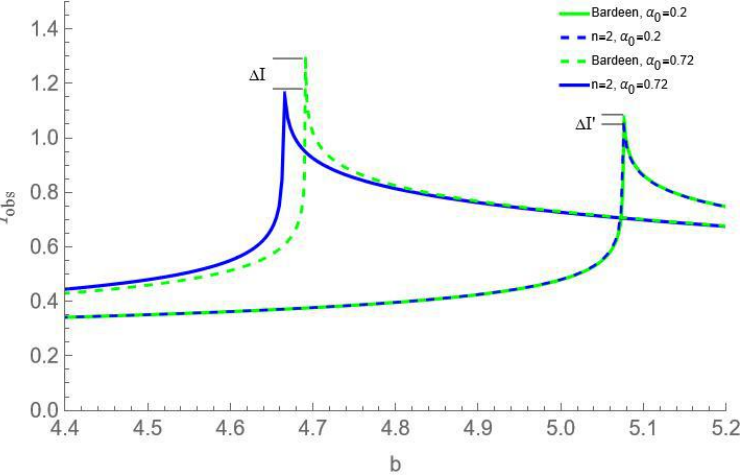}
        \label{fig:7a}
    }
    \hfill
    \subfloat[$n=2$, $\alpha_0=0.72$]{%
        \includegraphics[width=0.28\textwidth]{f3_d.pdf}
        \label{fig:7b}
    }
    \hfill
    \subfloat[Bardeen, $\alpha_0=0.72$]{%
        \includegraphics[width=0.28\textwidth]{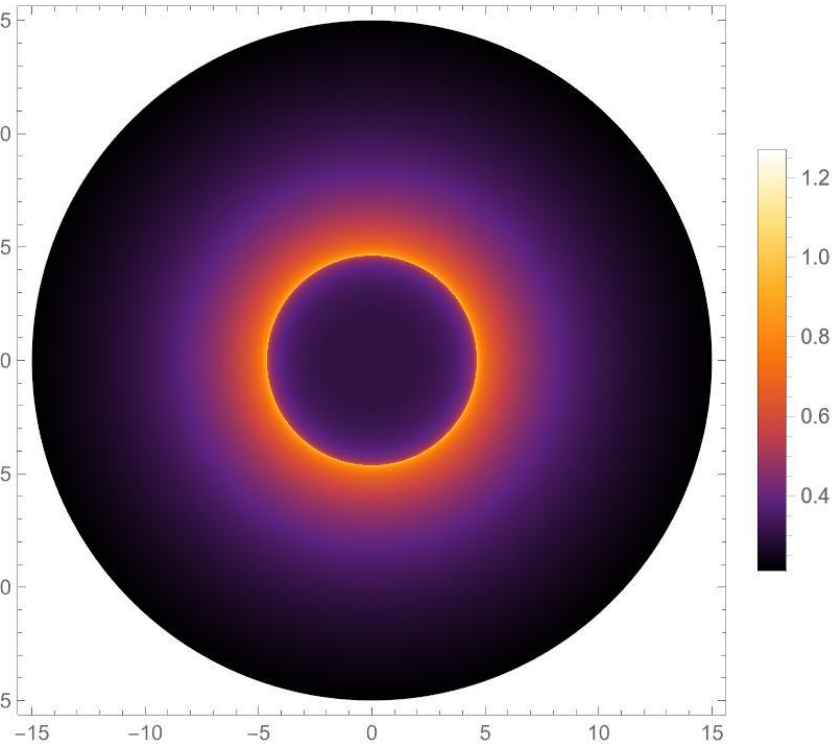}
        \label{fig:7c}
    }
    \caption{ The observed specific intensity and the optical appearances under the static spherical accretion for the new regular BH with $n=2$ and the corresponding Bardeen BH.}
    \label{fig:7}
\end{figure}

\begin{figure}[H]
    \centering
    \subfloat[fixing $n=3$ and changing $\alpha_0$]{%
        \includegraphics[width=0.38\textwidth]{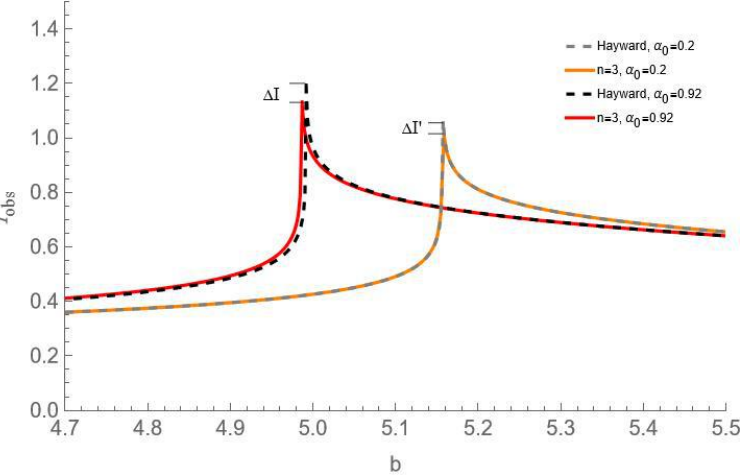}
        \label{fig:8a}
    }
    \hfill
    \subfloat[$n=3$, $\alpha_0=0.92$]{%
        \includegraphics[width=0.28\textwidth]{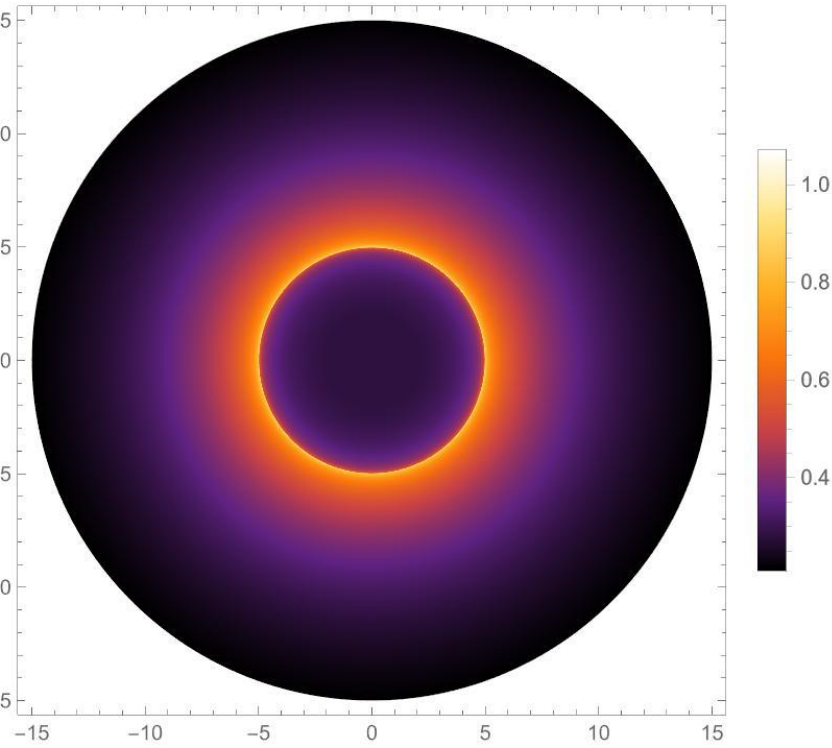}
        \label{fig:8b}
    }
    \hfill
    \subfloat[Hayward, $\alpha_0=0.92$]{%
        \includegraphics[width=0.28\textwidth]{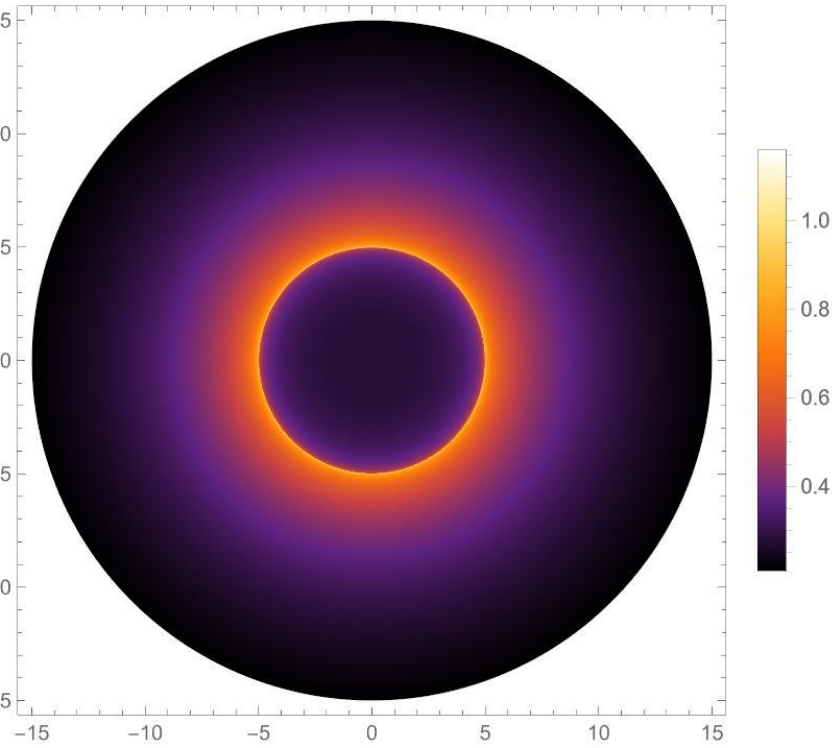}
        \label{fig:8c}
    }
    \caption{ The observed specific intensity and the optical appearances under static spherical accretion for the new regular BH with $n=3$ and the corresponding Hayward BH.}
    \label{fig:8}
\end{figure}

According to the previous discussion, when $n = 2$, $x = 2/3$, and when $n = 3$, $x = 1$, these new regular BHs show a one-to-one correspondence with the Bardeen BH and the Hayward BH respectively on a large scale. It is noted that the cores of these new regular BHs and the traditional regular BHs are different at the center. To compare the shadows of regular BHs with different cores, Figs.\ref{fig:7} and \ref{fig:8} present the observed specific intensity and optical appearances of the two types of regular BHs with different cores under the static accretion.

\end{multicols}
\begin{multicols}{2}

\end{multicols}
\begin{multicols}{2}
Firstly, from Figs.(\ref{fig:7a}) and (\ref{fig:8a}), it can be observed that the peak intensity of these new regular BHs with a Minkowski core is always smaller than that of the traditional regular BHs with a dS core. Here, $\Delta I$ and $\Delta I'$ represent the differences in the peak intensities of these regular BHs with different cores at the same $\alpha_0$ respectively, and $\Delta I > \Delta I'$. This implies that the larger $\alpha_0$ is, the more obvious the difference in the peak intensities of these BHs with different cores becomes, and the more significant the corresponding difference in the total observed intensity is.

Secondly, Figs.(\ref{fig:7b})(\ref{fig:7c})(\ref{fig:8b}) and (\ref{fig:8c}) present the optical appearances of these BHs with different cores when $\alpha_0=0.72$ and $\alpha_0=0.92$ respectively. It can be clearly seen that the photon ring of these regular BHs with a Minkowski core is significantly dimmer than that of the traditional regular BHs with a dS core. Moreover, the shadow radius and photon sphere radius of the former are also smaller than those of the latter. This indicates that, under the static spherical accretion model, compared with the traditional regular BHs, these new regular BHs have a dimmer photon ring, and the two exhibit different optical observational characteristics. as $\alpha_0$ increases, the differences in the optical observational features between these regular BHs with different cores become more significant, making it easier to distinguish them from their optical features.

Finally, by comparing Figs.(\ref{fig:7}) and (\ref{fig:8}), it can be seen that under the static spherical accretion, the difference between the Bardeen BH and the new regular BH corresponding to $n=2$ is more obvious than that between the Hayward BH and the new regular BH corresponding to $n=3$. This once again proves that the smaller $n$ is, the easier it is for observers to distinguish these regular BHs with different cores.

\end{multicols}
\begin{multicols}{2}

\subsection{Comparison under the infalling spherical accretion} \label{sec4-2}

\begin{figure}[H]
    \centering
    \subfloat[fixing $n=2$ and changing $\alpha_0$]{%
        \raisebox{-0.5\height}{\includegraphics[width=0.38\textwidth]{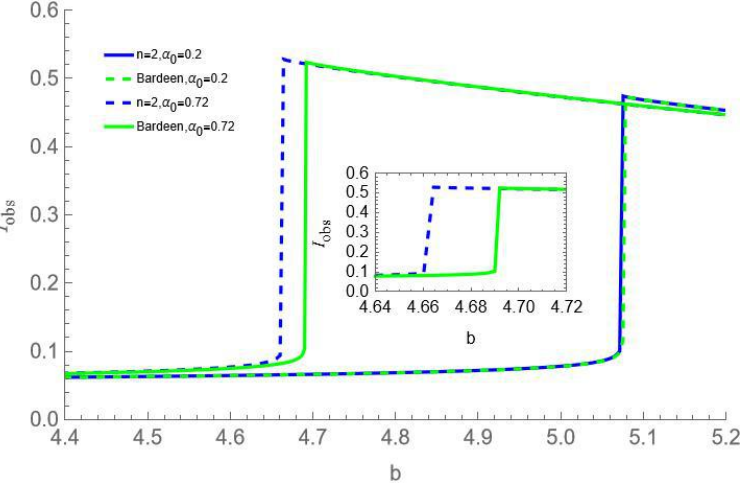}}
        \label{fig:9a}
    }
    \hfill
    \subfloat[$n=2$, $\alpha_0=0.72$]{%
        \raisebox{-0.5\height}{\includegraphics[width=0.25\textwidth]{f7_d.pdf}}
        \label{fig:9b}
    }
    \hfill
    \subfloat[Bardeen, $\alpha_0=0.72$]{%
        \raisebox{-0.5\height}{\includegraphics[width=0.25\textwidth]{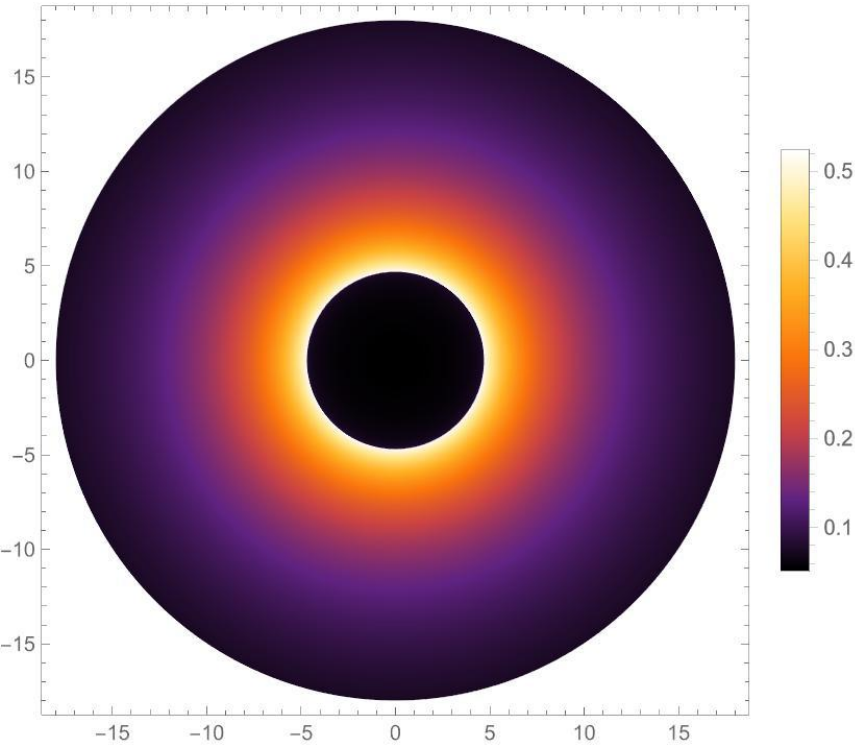}}
        \label{fig:9c}
    }
    \caption{ The observed specific intensity and the optical appearances under the infalling spherical accretion for the new regular BH with $n=2$ and the corresponding Bardeen BH.}
    \label{fig:9}
\end{figure}

\begin{figure}[H]
   \centering
   \subfloat[fixing $n=3$ and changing $\alpha_0$]{%
       \raisebox{-0.5\height}{\includegraphics[width=0.38\textwidth]{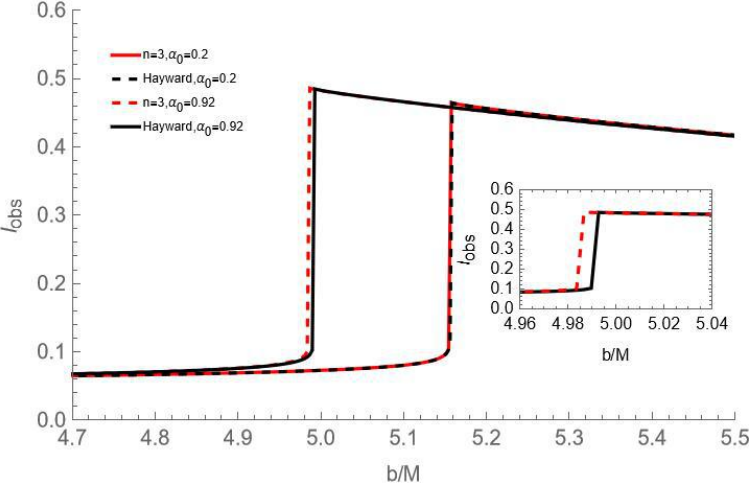}}
       \label{fig:10a}
   }
   \hfill
   \subfloat[$n=3$, $\alpha_0=0.92$]{%
       \raisebox{-0.5\height}{\includegraphics[width=0.25\textwidth]{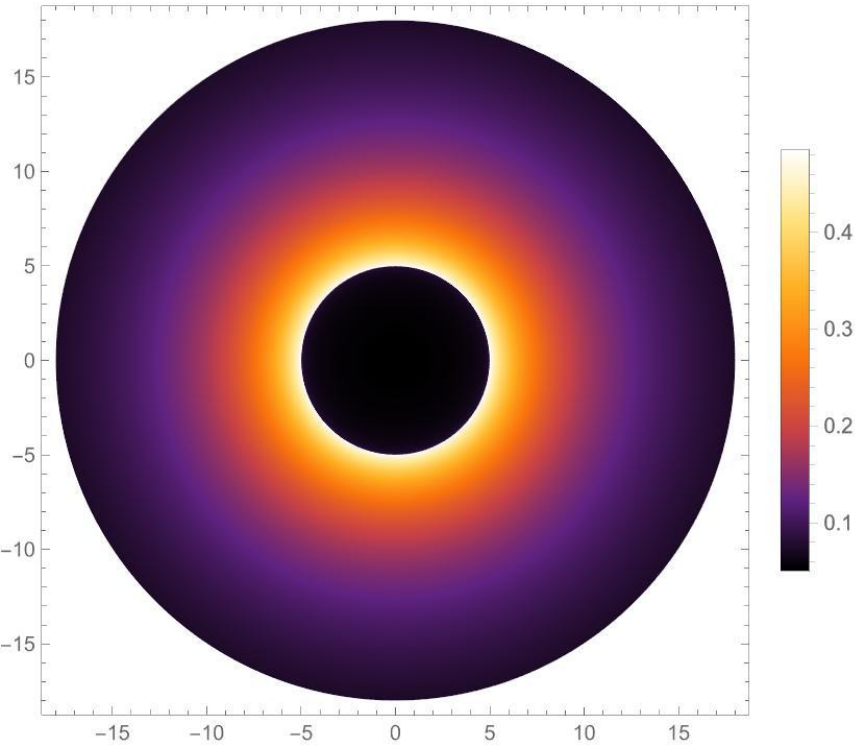}}
       \label{fig:10b}
   }
   \hfill
   \subfloat[Hayward, $\alpha_0=0.92$]{%
       \raisebox{-0.5\height}{\includegraphics[width=0.25\textwidth]{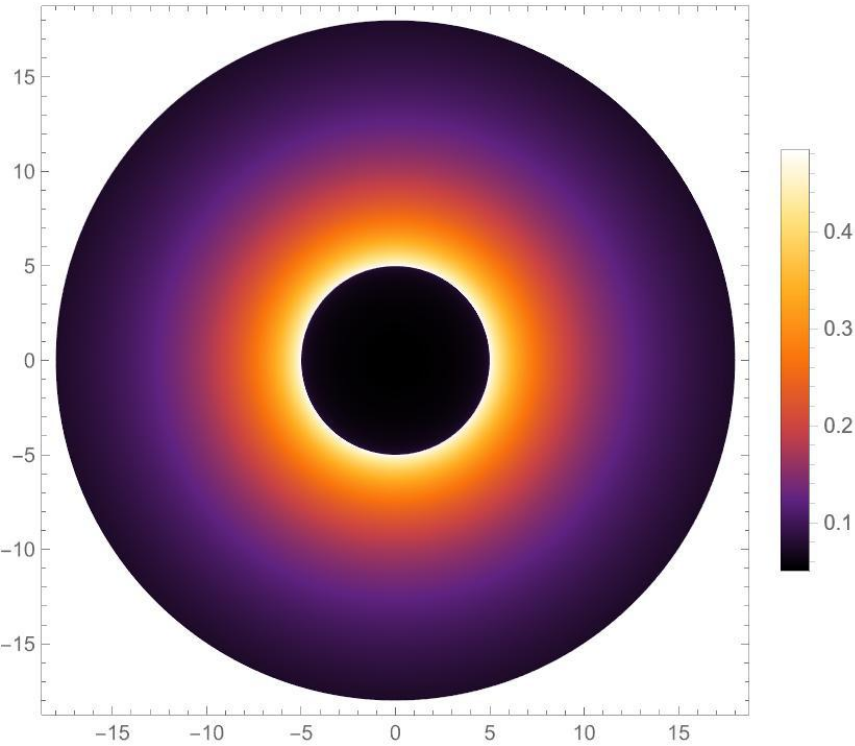}}
       \label{fig:10c}
   }
   \caption{ The observed specific intensity and the optical appearances under the infalling spherical accretion for the new regular BH with $n=3$ and the corresponding Hayward BH.}
   \label{fig:10}
\end{figure}

Under the infalling spherical accretion model, the observed specific intensity and optical appearances of these new regular BHs with $n=2$ and $n=3$, and their corresponding Bardeen/Hayward BHs, are shown in Figs.\ref{fig:9} and \ref{fig:10} respectively. It can be observed that, there are only slight differences in the shadows between the regular BHs with a Minkowski core and the traditional regular BHs with a dS core. As $\alpha_0$ increases, the shadow radius and the photon sphere radius of the former are slightly smaller than those of the latter, and the photon ring of the former is slightly dimmer than that of the latter. This difference becomes slightly more obvious only when $\alpha_0$ is relatively large.

\end{multicols}
\begin{multicols}{2}

\end{multicols}
\begin{multicols}{2}
The shadows of these new regular BHs with a Minkowski core and the corresponding traditional regular BHs with a dS core (Bardeen/Hayward BHs) are analyzed comparatively under two spherical accretion models. The shadows of these new regular BHs and the traditional regular BHs differ. Specifically, these new regular BHs have a smaller peak intensity, a dimmer photon ring, as well as smaller radius of BH shadow and photon sphere. Moreover, as $\alpha_0$ increases, these differences become more pronounced, making it easier to distinguish these regular BHs with different cores through optical observational characteristics. In addition, these differences are more pronounced under the static spherical accretion model.
\section{Conclusion} \label{sec5}

In this study, we explored the shadows of new regular BHs with a Minkowski core under different spherical accretion models and compared them with the traditional regular BHs with a de Sitter (dS) core (Bardeen/Hayward BHs). The results reveal that in both the static sphere accretion and the infalling sphere accretion, the observed specific intensity, shadow radius and photon sphere radius of these new regular BHs are closely associated with these intrinsic parameters. As the quantum gravity parameter $\alpha_0$ increases, the observed specific intensity increases, but the shadow radius and the photon sphere radius decrease. On the other hand, as the spacetime deformation parameter $n$ increases, the observed specific intensity decreases, while the shadow radius and the photon sphere radius increase. Moreover, the shadow radius and photon sphere radius are unaffected by the choice of the spherical accretion model, while the observed specific intensity differs between the models, with the higher observed specific intensity in the static sphere accretion than in the infalling sphere accretion.

Secondly, there are difference in the shadows of the two types of regular BHs with different cores, and these differences are more obvious in static spherical accretion. Taking the new regular BH with a Minkowski core($n=2$, $x=2/3$) and the Bardeen BH as an example, the former has a smaller observed specific intensity, a dimmer photon ring, smaller shadow radius and photon sphere radius. More importantly, these differences become more significant as $\alpha_0$ increases. Similarly, the new regular BH ($n=3$ and $x=1$) and the Hayward BH exhibit a similar trend. In the infalling sphere accretion, there are only subtle differences in observed specific intensity, and the differences are only slightly more obvious when $\alpha_0$ is relatively large.

Additionally, the results show that the shadows reflect the intrinsic properties of the BH spacetime, and the choice of the spherical accretion model only affects the observed specific intensity of the shadows and does not affect the structure of the BH shadow itself. Therefore, the spacetime characteristics of these new regular BHs can be distinguished by observing and studying the BH shadow.

In this study, we only apply two models: the static spherical accretion and the infalling spherical accretion. In fact, more complex accretion models can be studied, such as those incorporating the effects of magnetic fields or non-spherical accretion. It will help us to more comprehensively understand the changes in the shadows of BHs under different accretion scenarios and provide a more abundant theoretical foundation for astronomical observations. Meanwhile, our focus in this study was on comparing these new regular BHs with a Minkowski core to traditional regular BHs with a de Sitter core. Future research could expand the parameter ranges to study BH shadows under a wider variety of parameter combinations. With ongoing  advances in astronomical observation technology, the accuracy and scope of observational data continue to improve. Strengthening the integration of theoretical research with experimental observations is crucial. High resolution BH images and other observational data can be used to more precisely test theoretical models. By comparing  theoretical shadow calculations with observational results, models  can be continuously refined, enhancing our understanding of BH properties. Furthermore, studying the correlations between BH shadows and other astrophysical phenomena, such as BH spin and the distribution and dynamics of matter around BHs is important. Exploring how these factors jointly  influence the observational characteristics of shadows will contribute to a more comprehensive understanding of the physical processes around BHs and their role in galaxy evolution.

\acknowledgments{ This work is supported by Sichuan Science and Technology Program(2023NSFSC1352), and by the starting fund of China West Normal University (Nos. 20E069, 20A013 and 22kA005).}

\end{multicols}

\vspace{10mm}

\vspace{-1mm}
\centerline{\rule{80mm}{0.1pt}}
\vspace{2mm}

\begin{multicols}{2}

\end{multicols}

\clearpage

\end{document}